\documentclass[conference]{IEEEtran}
\IEEEoverridecommandlockouts

\usepackage{amsmath,amssymb,amsfonts}
\usepackage{algorithm}
\usepackage{algorithmicx}
\usepackage{algpseudocode}
\usepackage{setspace}
\usepackage{bm}
\usepackage{graphicx}
\usepackage{textcomp}
\usepackage{xcolor}
\usepackage{tikz}
\usetikzlibrary{quantikz2}
\usepackage{float}
\usepackage{subcaption}
\usepackage[T1]{fontenc}
\usepackage{verbatim}
\usepackage{hyperref}
\usepackage{makecell}
\usepackage{booktabs}
\usepackage{array} 
\usepackage{amsthm}
\usepackage{graphbox} 
\usepackage{adjustbox}
\usepackage{balance}
\usepackage{multirow}
\usepackage[capitalize]{cleveref}
\usepackage{footnote}
\usepackage{bm}
\usepackage{threeparttable}
\usepackage{caption}
\usepackage{lipsum}

\def\BibTeX{{\rm B\kern-.05em{\sc i\kern-.025em b}\kern-.08em
    T\kern-.1667em\lower.7ex\hbox{E}\kern-.125emX}}

\newtheorem{theo}{Theorem}
\newtheorem{lem}{Lemma}
\newtheorem{cor}{Corollary}
\theoremstyle{remark}

\theoremstyle{definition}
\newtheorem{defin}{Definition}

\newtheorem*{remark}{Remark}

\newtheorem*{prob}{Research Problem}

\newcommand{\eqdef}{\stackrel{\triangle}{=}}

\begin{document}

\title{Polynomial-time Extraction \\of Entanglement Resources
}

\author{Si-Yi Chen, Angela~Sara~Cacciapuoti,~\IEEEmembership{Senior~Member,~IEEE}, Marcello~Caleffi,~\IEEEmembership{Senior~Member,~IEEE}
    \thanks{This work has been accepted in the Proceedings of IEEE QCE’25.}
    \thanks{The authors are with the Quantum Internet Research Group \href{www.quantuminternet.it}{(www.QuantumInternet.it)}, University of Naples Federico II, Naples, 80125 Italy. }    
    \thanks{Corresponding author: Angela Sara Cacciapuoti (e-mail: angelasara.cacciapuoti@unina.it).}
    \thanks{This work has been funded by the European Union under Horizon Europe ERC-CoG grant QNattyNet, n.101169850. Views and opinions expressed are however those of the author(s) only and do not necessarily reflect those of the European Union or the European Research Council Executive Agency. Neither the European Union nor the granting authority can be held responsible for them.}
    }

\maketitle

\begin{abstract}
The extraction of EPR pairs and $n$-qubits GHZ states among remote nodes in quantum networks constitutes the resource primitives for end-to-end and on-demand communications. However, the Bell-VM problem, which determines whether a given graph state can be transformed into a set of Bell pairs on specific vertices (not necessarily remote), is known to be NP-complete. In this paper, we extend this problem, not only by focusing on nodes remote within generic graph states, but also by determining the number of extractable $n$-qubit remote GHZ states -- beside the number of remote EPR pairs. The rationale for tackling  the extraction of GHZ states among remote nodes, rather than solely remote EPR pairs, is that a GHZ state enables the dynamic extraction of an EPR pair between any pair of nodes sharing the state. This, in turn, implies the ability of accommodating the traffic requests on-the-fly. Specially, we propose a polynomial-time algorithm for solving the aforementioned NP-complete problem. Our results demonstrate that the proposed algorithm is able to effectively adapt to generic graph states for extracting entanglement resources across remote nodes.
\end{abstract}

\begin{IEEEkeywords}
Multipartite Entanglement, Entanglement-enabled connectivity, Quantum Networks, Quantum Internet, ERC-CoG QNattyNet.
\end{IEEEkeywords}

\section{Introduction}
\label{sec:1}

Entanglement is unanimously recognized as the key communication resource of the Quantum Internet. In particular, multipartite entanglement, i.e.,entanglement shared among more than two parties, enables a novel form of connectivity termed entanglement-enabled connectivity \cite{IllCalMan-22}. This connectivity, with no-counterpart in classical networks, activates a  flexible and dynamic overlay topology, referred to as \textit{artificial topology} \cite{IllCalMan-22,CheIllCac-25,PirDur-18}, upon the physical topology.

\begin{figure}[t]
    \centering
    \small
    \begin{subfigure}[b]{1\columnwidth}
        \centering
        \includegraphics[width=0.221\columnwidth]{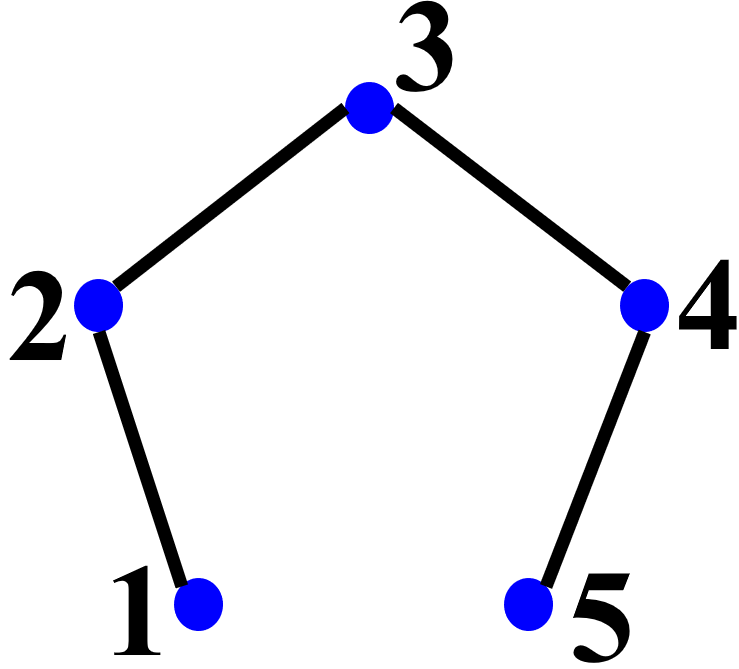}
        \subcaption{Initial artificial topology: 5-qubits linear graph state}
        \label{fig:01.a}
    \end{subfigure}
    
    \vspace{6pt}
    
    \begin{subfigure}[b]{\columnwidth}
        \begin{subfigure}[t]{0.47\columnwidth}
            \centering
            \includegraphics[width=0.47\columnwidth]{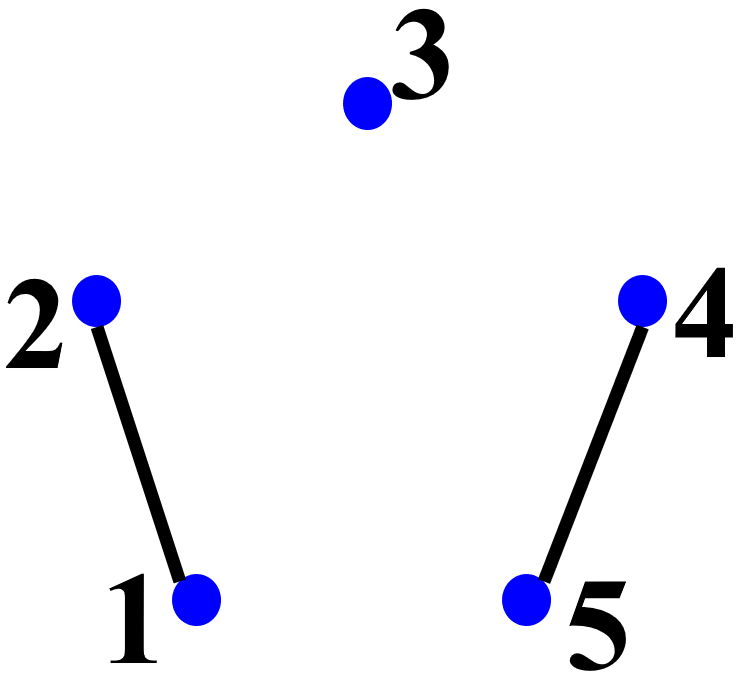}
            \subcaption{Vanilla Pairability: two EPR pairs can be extracted from the artificial topology of Fig.~\ref{fig:01.a}.}
            \label{fig:01.b}
        \end{subfigure}
        \hspace{6pt}
        \begin{subfigure}[t]{0.47\columnwidth}
            \centering
            \includegraphics[width=0.47\columnwidth]{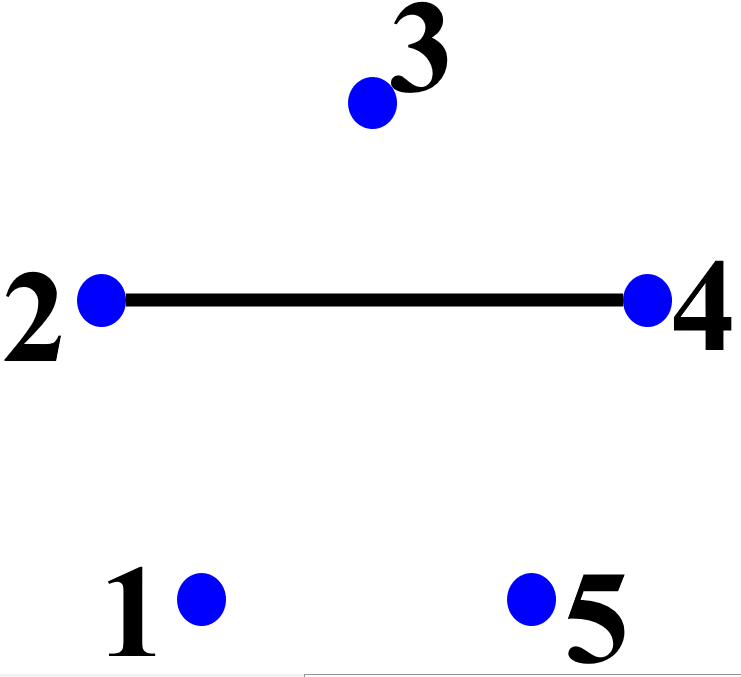}
            \subcaption{Remote Pairability: only one EPR pair between remote nodes can be extracted from the artificial topology of Fig.~\ref{fig:01.a}.}
            \label{fig:01.c}
        \end{subfigure}
    \end{subfigure}

    \vspace{6pt}
    
    \begin{subfigure}[b]{\columnwidth} 
        \begin{subfigure}[t]{0.47\columnwidth}
            \centering
            \includegraphics[width=0.47\columnwidth]{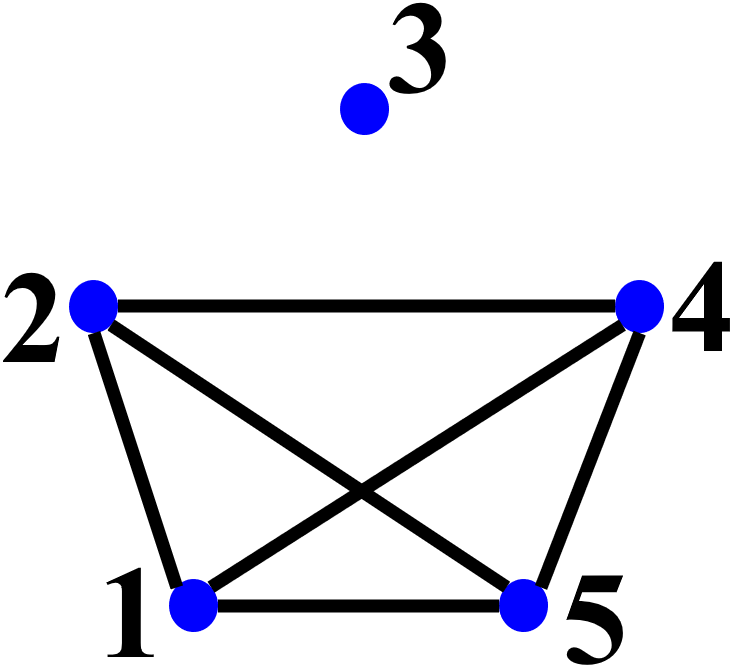}
            \subcaption{Vanilla Gability: one $4$-qubit GHZ state can be extracted from the artificial topology of Fig.~\ref{fig:01.a}.}
            \label{fig:01.d}
        \end{subfigure}
        \hspace{6pt}
        \begin{subfigure}[t]{0.47\columnwidth}
            \centering
            \includegraphics[width=0.47\columnwidth]{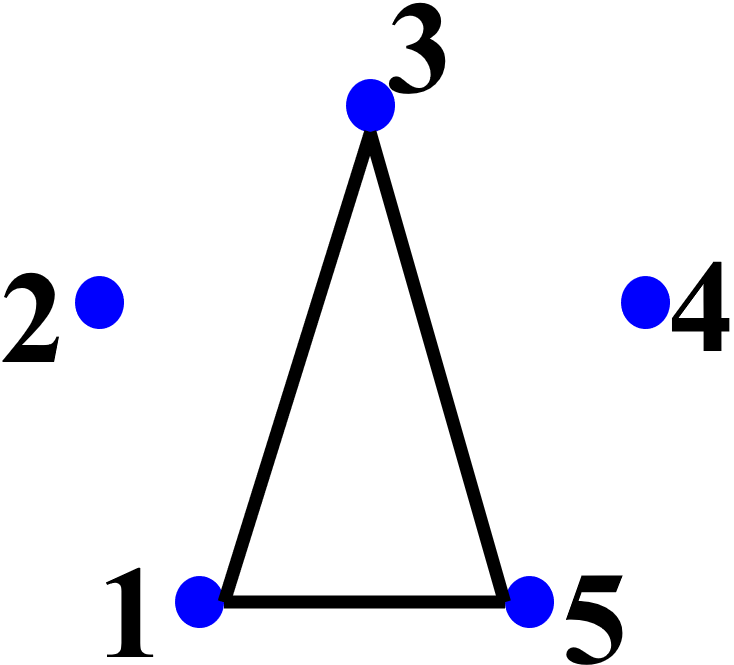}
            \subcaption{Remote Gability: one $3$-qubit GHZ state between remote nodes can be extracted from the artificial topology of Fig.~\ref{fig:01.a}.}
            \label{fig:01.e}
        \end{subfigure}
    \end{subfigure}
    \caption{\textit{Remote} vs \textit{vanilla} Pairability and Gability for a 5-qubit linear graph state. Fig.~\ref{fig:01.a} shows the initial artificial topology enabled by a 5-qubits linear graph state. When it comes to Pairability, up to $2$ EPR pairs can be extracted from the considered graph state. Yet, the extracted EPRs in Fig.~\ref{fig:01.b} ``\textit{link}'' nodes that are already connected in the initial artificial topology. Conversely, only one remote EPR in Fig.~\ref{fig:01.c} can be extracted in the same artificial topology. When it comes to $n$-Gability, the differences are even more crucial. Specifically, GHZ state with mass up to $4$ qubits as shown in Fig.~\ref{fig:01.d} can be extracted among the nodes, regardless of their proximity in the artificial topology. Conversely, if we constraint the nodes to be remote as in Def.~\ref{def:02}, the largest GHZ state has mass equal to $3$, as shown in Fig.~\ref{fig:01.e}.}
    \label{fig:01}
    \hrulefill
\end{figure}

As recently proved in \cite{CheIllCac-25, MazCalCac-24}, artificial topologies can be properly manipulated to account for the dynamic nature of the node communication requests. This has led to the very growing interest in studying the extraction of entanglement resources for communication purpose.

A representative example is the Bell-VM problem, which is known to be NP-complete \cite{DahHelWeh-18}. Bell-VM asks whether a given graph state can be transformed into a set of $k$ Bell pairs between specific node pairs, by using only local Clifford operations, Pauli measurements, and classical communication (LC + LPM + CC). Research in this area has potential applications in point-to-point quantum communication protocols, where the extracted $k$ Bell pairs can be used for the parallel transmission of $k$ informational qubits via teleporting protocol \cite{CacCalVan-20}.  However, the locations of the $k$ node pairs must be chosen in advance, limiting the adaptability to the dynamics of the communication requests.

A related line of research focuses on determining whether a GHZ state can be extracted from a given graph state, by using only LC + LPM + CC operations. This problem has also been proven to be NP-complete~\cite{DahHelWeh-20}. For ease of reference, we refer to the aforementioned problem as \textit{GHZ-VM problem}. Although the GHZ-VM problem concerns the extraction of a single GHZ state, it offers a significant advantage over the Bell-VM formulation, since it inherently supports adaptability. Indeed, a GHZ state allows the dynamic extraction of an EPR pair between any pair of nodes sharing the original GHZ state. And, notably, this extraction can occur at runtime, driven by the actual and potentially time-varying communication needs of the nodes.

It should be emphasized that the scope of both Bell-VM and GHZ-VM studies lies within the so-called vanilla extractions, as represented in Fig.~\ref{fig:01.b} and~\ref{fig:01.d}, taken from \cite{CheCacCal-25}. In other words, the extraction is performed regardless of the nodes proximity within the artificial topology. However, a more urgent and critical aspect lies in understanding the capabilities of extracting entanglement resources among nodes remote within the aritficial topology. In fact, having a fully-connected artificial topology among all the network nodes is not reasonable, due the challenges related to (and the complex equipment necessary for) the generation and the control of a complex multipartite state. Hence, it is more practical and logical to assume the presence of nodes that are remote in the artificial topology. And as highligted in  \cite{CheCacCal-25}, the extraction of EPR pairs and GHZ
states among remote nodes constitutes the
resource primitives for end-to-end and on-demand communications. 

\begin{figure}[t]
    \centering    \includegraphics[width=0.85\columnwidth]{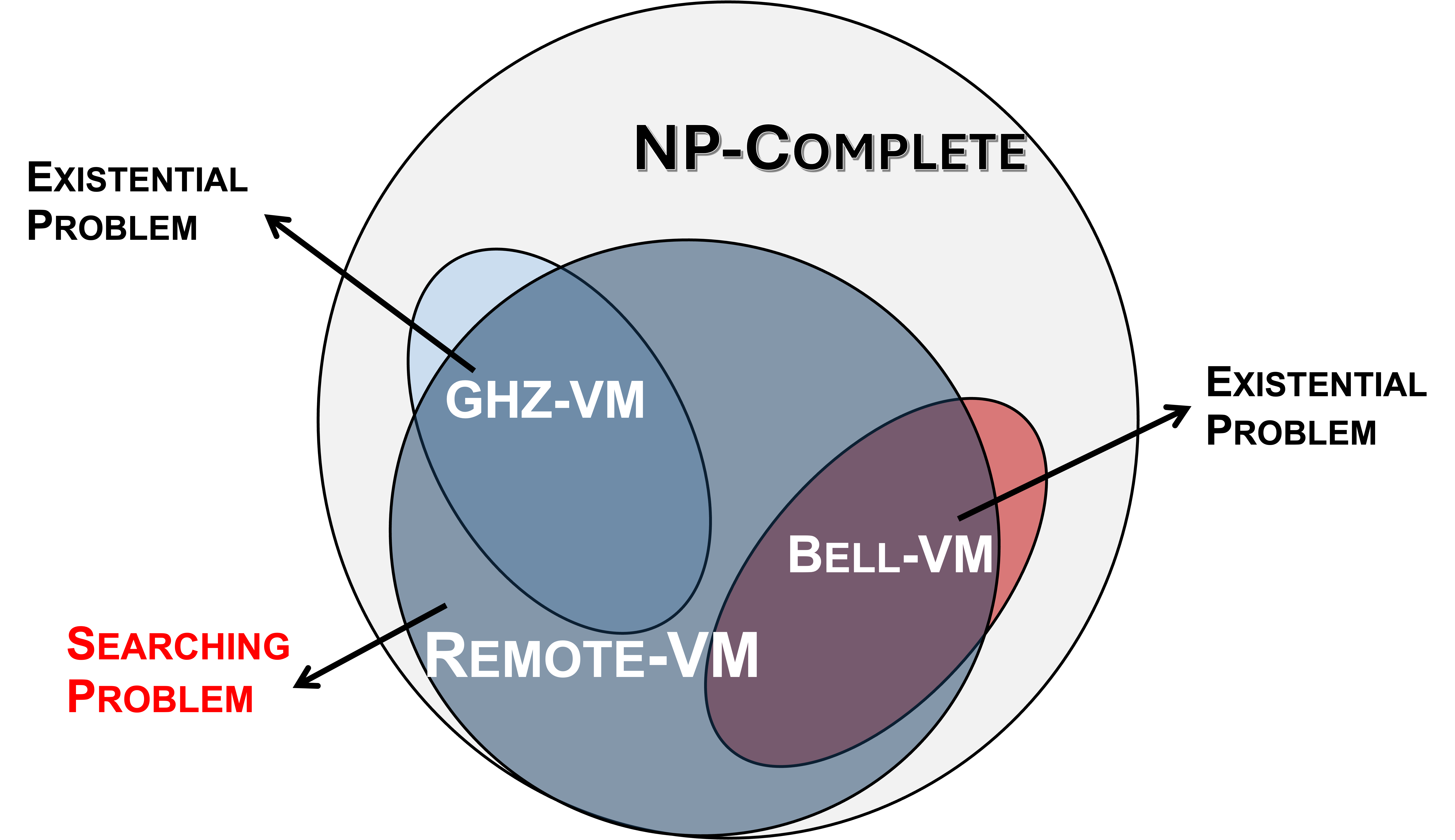}
    \caption{Venn diagram for the relationship between GHZ-VM, BELL-VM, and REMOTE-VM (our research problem). }
    \label{fig:x02}
    \hrulefill
\end{figure}

Accordingly, in this paper, we  extend the Bell-VM and GHZ-VM approaches to a deeper problem, referred to as \textbf{Remote-VM problem}, by determining the number of $n$-qubit GHZ states and Bell pairs that can be extracted among remote nodes of a given graph state, by using only single-qubit Clifford operations, single-qubit Pauli measurements, and classical communication. And, remarkably, we solve the NP-complete \textit{Remote-VM problem} through a polynomial-time algorithm. To the best of our knowledge, this is the first paper solving in polynomial-time this problem.\\
The difference between the Remote-VM and existing Bell-VM and GHZ-VM is pictorially represented in Fig.~\ref{fig:x02}. Specifically, while both Bell-VM and GHZ-VM are existential problems, the Remote-VM problem is a search problem, which answers not only to the question about the existence, but also determines the exact extractable resources. We summarize our contributions as follows:
\begin{itemize} 
    \item We propose a polynomial-time algorithm for the Remote-VM (see Sec.~\ref{sec:3.2}), by exploiting graph-theory tools and only LC + LPM + CC. 
    \item The algorithm facilitates the extraction of both $n$-qubit GHZ states and EPR pairs among remote nodes, by extending the extraction capabilities beyond a set of apriori selected EPR pairs or a specific GHZ state. 
    \item The algorithm is not restricted to any particular type of two-colorable graph, thereby broadening its applicability to general bipartite graphs (see Sec.~\ref{sec:4}). 
\end{itemize}



\section{Research Problem}
\label{sec:2}
\subsection{Preliminaries}
\label{sec:2.1}
As aforementioned, we focus on two-colorable graph states, which can be effectively described with graph theory tools. We assume that each qubit of the graph state is distributed to each network node, and we equivalently refer to node $i$ or to vertex $v_i$ associated with the qubit of the graph state $\ket{G}$ stored at such a node. To formally define our research problem, the following definitions are preliminary. For a more comprehensive background, we refer interested readers to~\cite{CheCacCal-25}.

\begin{defin}[\textbf{Remote Nodes}]
    \label{def:01}
    Given a $N$-qubit graph state $\ket{G}$ and its corresponding graph $G=(V,E)$, two network nodes $i$ and $j$ are defined as \textit{remote} if the corresponding vertices $v_i,v_j$ are non-adjacent in $G$, i.e., if:
    \begin{equation}
        \label{eq:01}
        (v_i,v_j) \not\in E.
    \end{equation}
\end{defin}


\begin{defin}[\textbf{$\bm{{r_g}(n)}$: remote $\bf{n}$-Gability}]
    \label{def:02}
    The remote $n$-Gability of an $N$-qubit graph state $\ket{G}$ quantifies the number of $n$-qubit GHZ states ($n \leq N$) that can be extracted among remote nodes using LOCC. The number of such GHZ states extractable from $\ket{G}$ is denoted by $r_g(n)$.
\end{defin}

\begin{remark}
    Since an EPR pair is a special case of a GHZ state with two qubits, the case of $r_g(2)$ is essentially a special instance of remote $n$-Gability. We refer to this as remote Pairability, which formal defined in Def.~\ref{def:03}.
\end{remark}

\begin{defin}[\textbf{$\bm{{r_g}(2)}$: remote Pairability}]
    \label{def:03}
    The remote Pairability of an $N$-qubit graph state $\ket{G}$ quantifies the number of EPR pairs that can be extracted among remote nodes using LOCC. The number of extractable EPR pairs from a graph state $\ket{G}$ is denoted by $r_g(2)$.
\end{defin}

\begin{defin}[\textbf{Two-colorable Graph} or \textbf{Bipartite Graph}]
    \label{def:04}
    A graph $G=(V,E)$ is two-colorable if the set of vertices $V$ can be partitioned into two subsets $\{ P_1, P_2 \}$ so that there exist no edge in $E$ between two vertices belonging to the same subset. Two-colorable graph $G=(V,E)$ can be also denoted as $G=(P_1, P_2, E)$.
\end{defin}

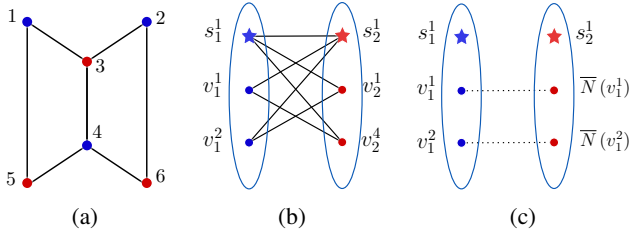
\begin{figure}[t]
    \centering
    \begin{minipage}[b]{0.5\textwidth}
    \centering
        \begin{minipage}[b]{0.28\textwidth}
            \centering\resizebox{\textwidth}{!}{
        \tikzset{every picture/.style={line width=0.75pt}} 

\begin{tikzpicture}[x=0.75pt,y=0.75pt,yscale=-1,xscale=1]

\draw   (490.67,50.48) -- (531.67,77.82) -- (531.67,135.32) -- (490.67,162.65) -- cycle ;
\draw  [color={rgb, 255:red, 255; green, 255; blue, 255 }  ,draw opacity=0.44 ][fill={rgb, 255:red, 12; green, 2; blue, 208 }  ,fill opacity=1 ] (486.77,50.48) .. controls (486.77,48.33) and (488.51,46.58) .. (490.67,46.58) .. controls (492.82,46.58) and (494.57,48.33) .. (494.57,50.48) .. controls (494.57,52.64) and (492.82,54.38) .. (490.67,54.38) .. controls (488.51,54.38) and (486.77,52.64) .. (486.77,50.48) -- cycle ;
\draw   (572.67,162.65) -- (531.67,135.32) -- (531.67,77.82) -- (572.67,50.48) -- cycle ;
\draw  [color={rgb, 255:red, 255; green, 255; blue, 255 }  ,draw opacity=0.44 ][fill={rgb, 255:red, 12; green, 2; blue, 208 }  ,fill opacity=1 ] (568.77,50.48) .. controls (568.77,48.33) and (570.51,46.58) .. (572.67,46.58) .. controls (574.82,46.58) and (576.57,48.33) .. (576.57,50.48) .. controls (576.57,52.64) and (574.82,54.38) .. (572.67,54.38) .. controls (570.51,54.38) and (568.77,52.64) .. (568.77,50.48) -- cycle ;
\draw  [color={rgb, 255:red, 255; green, 255; blue, 255 }  ,draw opacity=0.44 ][fill={rgb, 255:red, 12; green, 2; blue, 208 }  ,fill opacity=1 ] (527.77,135.32) .. controls (527.77,133.16) and (529.51,131.42) .. (531.67,131.42) .. controls (533.82,131.42) and (535.57,133.16) .. (535.57,135.32) .. controls (535.57,137.47) and (533.82,139.22) .. (531.67,139.22) .. controls (529.51,139.22) and (527.77,137.47) .. (527.77,135.32) -- cycle ;
\draw  [color={rgb, 255:red, 255; green, 255; blue, 255 }  ,draw opacity=0.44 ][fill={rgb, 255:red, 208; green, 2; blue, 2 }  ,fill opacity=1 ] (527.77,77.82) .. controls (527.77,75.66) and (529.51,73.92) .. (531.67,73.92) .. controls (533.82,73.92) and (535.57,75.66) .. (535.57,77.82) .. controls (535.57,79.97) and (533.82,81.72) .. (531.67,81.72) .. controls (529.51,81.72) and (527.77,79.97) .. (527.77,77.82) -- cycle ;
\draw  [color={rgb, 255:red, 255; green, 255; blue, 255 }  ,draw opacity=0.44 ][fill={rgb, 255:red, 208; green, 2; blue, 2 }  ,fill opacity=1 ] (486.77,161.15) .. controls (486.77,158.99) and (488.51,157.25) .. (490.67,157.25) .. controls (492.82,157.25) and (494.57,158.99) .. (494.57,161.15) .. controls (494.57,163.3) and (492.82,165.05) .. (490.67,165.05) .. controls (488.51,165.05) and (486.77,163.3) .. (486.77,161.15) -- cycle ;
\draw  [color={rgb, 255:red, 255; green, 255; blue, 255 }  ,draw opacity=0.44 ][fill={rgb, 255:red, 208; green, 2; blue, 2 }  ,fill opacity=1 ] (568.77,161.15) .. controls (568.77,158.99) and (570.51,157.25) .. (572.67,157.25) .. controls (574.82,157.25) and (576.57,158.99) .. (576.57,161.15) .. controls (576.57,163.3) and (574.82,165.05) .. (572.67,165.05) .. controls (570.51,165.05) and (568.77,163.3) .. (568.77,161.15) -- cycle ;

\draw (475.67,40.42) node [anchor=north west][inner sep=0.75pt]    {$1$};
\draw (577.33,41.75) node [anchor=north west][inner sep=0.75pt]    {$2$};
\draw (536,75) node [anchor=north west][inner sep=0.75pt]    {$3$};
\draw (534,120.08) node [anchor=north west][inner sep=0.75pt]    {$4$};
\draw (474.67,152.42) node [anchor=north west][inner sep=0.75pt]    {$5$};
\draw (577.33,150.42) node [anchor=north west][inner sep=0.75pt]    {$6$};

\end{tikzpicture}
    }
            \subcaption{}
    	   \label{fig:02.1}
        \end{minipage}
        \begin{minipage}[b]{0.3\textwidth}
            \centering\resizebox{\textwidth}{!}{
        \tikzset{every picture/.style={line width=0.75pt}} 

\begin{tikzpicture}[x=0.75pt,y=0.75pt,yscale=-1,xscale=1]

\draw    (80.3,58.99) -- (160.3,58.66) ;
\draw    (80.3,58.99) -- (160.1,105.04) ;
\draw    (80.3,58.99) -- (160.1,150.04) ;
\draw    (80.1,150.38) -- (160.3,58.66) ;
\draw    (80.1,105.38) -- (160.3,58.66) ;
\draw    (80.1,105.38) -- (160.1,150.04) ;
\draw    (80.1,150.38) -- (160.1,105.04) ;
\draw  [color={rgb, 255:red, 12; green, 90; blue, 182 }  ,draw opacity=1 ] (63,110.28) .. controls (63,66.15) and (70.66,30.38) .. (80.1,30.38) .. controls (89.54,30.38) and (97.2,66.15) .. (97.2,110.28) .. controls (97.2,154.41) and (89.54,190.18) .. (80.1,190.18) .. controls (70.66,190.18) and (63,154.41) .. (63,110.28) -- cycle ;
\draw  [color={rgb, 255:red, 255; green, 255; blue, 255 }  ,draw opacity=0.38 ][fill={rgb, 255:red, 54; green, 58; blue, 230 }  ,fill opacity=1 ][line width=0.75]  (80.3,50.18) -- (82.68,55.42) -- (88,56.27) -- (84.15,60.35) -- (85.06,66.12) -- (80.3,63.39) -- (75.54,66.12) -- (76.45,60.35) -- (72.6,56.27) -- (77.92,55.42) -- cycle ;
\draw  [color={rgb, 255:red, 255; green, 255; blue, 255 }  ,draw opacity=0.44 ][fill={rgb, 255:red, 12; green, 2; blue, 208 }  ,fill opacity=1 ] (76.2,105.38) .. controls (76.2,103.22) and (77.95,101.48) .. (80.1,101.48) .. controls (82.25,101.48) and (84,103.22) .. (84,105.38) .. controls (84,107.53) and (82.25,109.28) .. (80.1,109.28) .. controls (77.95,109.28) and (76.2,107.53) .. (76.2,105.38) -- cycle ;
\draw  [color={rgb, 255:red, 255; green, 255; blue, 255 }  ,draw opacity=0.44 ][fill={rgb, 255:red, 12; green, 2; blue, 208 }  ,fill opacity=1 ] (76.2,150.38) .. controls (76.2,148.22) and (77.95,146.48) .. (80.1,146.48) .. controls (82.25,146.48) and (84,148.22) .. (84,150.38) .. controls (84,152.53) and (82.25,154.28) .. (80.1,154.28) .. controls (77.95,154.28) and (76.2,152.53) .. (76.2,150.38) -- cycle ;
\draw  [color={rgb, 255:red, 12; green, 90; blue, 182 }  ,draw opacity=1 ] (143,109.94) .. controls (143,65.82) and (150.66,30.04) .. (160.1,30.04) .. controls (169.54,30.04) and (177.2,65.82) .. (177.2,109.94) .. controls (177.2,154.07) and (169.54,189.84) .. (160.1,189.84) .. controls (150.66,189.84) and (143,154.07) .. (143,109.94) -- cycle ;
\draw  [color={rgb, 255:red, 255; green, 255; blue, 255 }  ,draw opacity=0.38 ][fill={rgb, 255:red, 230; green, 54; blue, 54 }  ,fill opacity=1 ][line width=0.75]  (160.3,49.84) -- (162.68,55.09) -- (168,55.93) -- (164.15,60.02) -- (165.06,65.78) -- (160.3,63.06) -- (155.54,65.78) -- (156.45,60.02) -- (152.6,55.93) -- (157.92,55.09) -- cycle ;
\draw  [color={rgb, 255:red, 255; green, 255; blue, 255 }  ,draw opacity=0.44 ][fill={rgb, 255:red, 208; green, 2; blue, 2 }  ,fill opacity=1 ] (156.2,105.04) .. controls (156.2,102.89) and (157.95,101.14) .. (160.1,101.14) .. controls (162.25,101.14) and (164,102.89) .. (164,105.04) .. controls (164,107.2) and (162.25,108.94) .. (160.1,108.94) .. controls (157.95,108.94) and (156.2,107.2) .. (156.2,105.04) -- cycle ;
\draw  [color={rgb, 255:red, 255; green, 255; blue, 255 }  ,draw opacity=0.44 ][fill={rgb, 255:red, 208; green, 2; blue, 2 }  ,fill opacity=1 ] (156.2,150.04) .. controls (156.2,147.89) and (157.95,146.14) .. (160.1,146.14) .. controls (162.25,146.14) and (164,147.89) .. (164,150.04) .. controls (164,152.2) and (162.25,153.94) .. (160.1,153.94) .. controls (157.95,153.94) and (156.2,152.2) .. (156.2,150.04) -- cycle ;

\draw (39,89.42) node [anchor=north west][inner sep=0.75pt]  [font=\Large]  {$v_{1}^{1}$};
\draw (39,136.42) node [anchor=north west][inner sep=0.75pt] [font=\Large]   {$v_{1}^{2}$};
\draw (40,42.42) node [anchor=north west][inner sep=0.75pt] [font=\Large]   {$s_{1}^{1}$};
\draw (176,89.42) node [anchor=north west][inner sep=0.75pt] [font=\Large]   {$v_{2}^{1}$};
\draw (175,136.42) node [anchor=north west][inner sep=0.75pt]  [font=\Large]  {$v_{2}^{4}$};
\draw (176,42.42) node [anchor=north west][inner sep=0.75pt] [font=\Large]    {$s_{2}^{1}$};

\end{tikzpicture}
    }
            \subcaption{}
    	   \label{fig:02.2}
        \end{minipage}
        \begin{minipage}[b]{0.35\textwidth}
            \centering\resizebox{\textwidth}{!}{
        \tikzset{every picture/.style={line width=0.75pt}} 

\begin{tikzpicture}[x=0.75pt,y=0.75pt,yscale=-1,xscale=1]

\draw  [dash pattern={on 0.84pt off 2.51pt}]  (272.1,105.38) -- (348.2,105.04) ;
\draw  [dash pattern={on 0.84pt off 2.51pt}]  (272.1,150.38) -- (348.2,150.04) ;
\draw  [color={rgb, 255:red, 12; green, 90; blue, 182 }  ,draw opacity=1 ] (255,110.28) .. controls (255,66.15) and (262.66,30.38) .. (272.1,30.38) .. controls (281.54,30.38) and (289.2,66.15) .. (289.2,110.28) .. controls (289.2,154.41) and (281.54,190.18) .. (272.1,190.18) .. controls (262.66,190.18) and (255,154.41) .. (255,110.28) -- cycle ;
\draw  [color={rgb, 255:red, 255; green, 255; blue, 255 }  ,draw opacity=0.38 ][fill={rgb, 255:red, 54; green, 58; blue, 230 }  ,fill opacity=1 ][line width=0.75]  (272.3,50.18) -- (274.68,55.42) -- (280,56.27) -- (276.15,60.35) -- (277.06,66.12) -- (272.3,63.39) -- (267.54,66.12) -- (268.45,60.35) -- (264.6,56.27) -- (269.92,55.42) -- cycle ;
\draw  [color={rgb, 255:red, 255; green, 255; blue, 255 }  ,draw opacity=0.44 ][fill={rgb, 255:red, 12; green, 2; blue, 208 }  ,fill opacity=1 ] (268.2,105.38) .. controls (268.2,103.22) and (269.95,101.48) .. (272.1,101.48) .. controls (274.25,101.48) and (276,103.22) .. (276,105.38) .. controls (276,107.53) and (274.25,109.28) .. (272.1,109.28) .. controls (269.95,109.28) and (268.2,107.53) .. (268.2,105.38) -- cycle ;
\draw  [color={rgb, 255:red, 255; green, 255; blue, 255 }  ,draw opacity=0.44 ][fill={rgb, 255:red, 12; green, 2; blue, 208 }  ,fill opacity=1 ] (268.2,150.38) .. controls (268.2,148.22) and (269.95,146.48) .. (272.1,146.48) .. controls (274.25,146.48) and (276,148.22) .. (276,150.38) .. controls (276,152.53) and (274.25,154.28) .. (272.1,154.28) .. controls (269.95,154.28) and (268.2,152.53) .. (268.2,150.38) -- cycle ;
\draw  [color={rgb, 255:red, 12; green, 90; blue, 182 }  ,draw opacity=1 ] (335,109.94) .. controls (335,65.82) and (342.66,30.04) .. (352.1,30.04) .. controls (361.54,30.04) and (369.2,65.82) .. (369.2,109.94) .. controls (369.2,154.07) and (361.54,189.84) .. (352.1,189.84) .. controls (342.66,189.84) and (335,154.07) .. (335,109.94) -- cycle ;
\draw  [color={rgb, 255:red, 255; green, 255; blue, 255 }  ,draw opacity=0.38 ][fill={rgb, 255:red, 230; green, 54; blue, 54 }  ,fill opacity=1 ][line width=0.75]  (352.3,49.84) -- (354.68,55.09) -- (360,55.93) -- (356.15,60.02) -- (357.06,65.78) -- (352.3,63.06) -- (347.54,65.78) -- (348.45,60.02) -- (344.6,55.93) -- (349.92,55.09) -- cycle ;
\draw  [color={rgb, 255:red, 255; green, 255; blue, 255 }  ,draw opacity=0.44 ][fill={rgb, 255:red, 208; green, 2; blue, 2 }  ,fill opacity=1 ] (348.2,105.04) .. controls (348.2,102.89) and (349.95,101.14) .. (352.1,101.14) .. controls (354.25,101.14) and (356,102.89) .. (356,105.04) .. controls (356,107.2) and (354.25,108.94) .. (352.1,108.94) .. controls (349.95,108.94) and (348.2,107.2) .. (348.2,105.04) -- cycle ;
\draw  [color={rgb, 255:red, 255; green, 255; blue, 255 }  ,draw opacity=0.44 ][fill={rgb, 255:red, 208; green, 2; blue, 2 }  ,fill opacity=1 ] (348.2,150.04) .. controls (348.2,147.89) and (349.95,146.14) .. (352.1,146.14) .. controls (354.25,146.14) and (356,147.89) .. (356,150.04) .. controls (356,152.2) and (354.25,153.94) .. (352.1,153.94) .. controls (349.95,153.94) and (348.2,152.2) .. (348.2,150.04) -- cycle ;

\draw (232,91.42) node [anchor=north west][inner sep=0.75pt]  [font=\Large]  {$v_{1}^{1}$};
\draw (232,137.42) node [anchor=north west][inner sep=0.75pt]  [font=\Large]  {$v_{1}^{2}$};
\draw (373,91.42) node [anchor=north west][inner sep=0.75pt]  [font=\large]  
{$\overline{N}\left( v_{1}^{1}\right)$};
\draw (373,137.42) node [anchor=north west][inner sep=0.75pt]  [font=\large]  
{$\overline{N}\left( v_{1}^{2}\right)$};
\draw (233,43.42) node [anchor=north west][inner sep=0.75pt]  [font=\Large]  {$s_{1}^{1}$};
\draw (369,43.42) node [anchor=north west][inner sep=0.75pt]  [font=\Large]  {$s_{2}^{1}$};

\end{tikzpicture}
    }
            \subcaption{}
    	   \label{fig:02.3}
        \end{minipage}
    \end{minipage}
    \caption{Pictorial representation for opposite remote sets. (a) Original butterfly graph. (b) Vertex partitioning, same topology as in Fig.\ref{fig:02.1}, with star vertices at nodes 3 and 4. (c) Opposite remote sets for vertices $v^1_1$ and $v^2_1$ in Fig.\ref{fig:02.2}; dashed lines indicate opposite remote sets per Def.~\ref{def:05}.} 
    \label{fig:02}
    \hrulefill
\end{figure}

\begin{defin}[\textbf{Opposite Remote-Set}]
    \label{def:05}
    Given a two-colorable graph $G=(P_1, P_2, E)$, the \textit{opposite remote set} of the arbitrary vertex $v_i \in P_i$, with $i \in \{1,2\}$, is the set $\overline{N}(v_i)$ of remote vertices of $v_i$ belonging to the other partition:
    \begin{equation}
        \label{eq:02}
        \overline{N}(v_i) \eqdef \big\{ v_j \in P_j \neq P_i : (v_i,v_j ) \not\in E \big\}.
    \end{equation}
\end{defin}

The term ``opposite'' in Def.~\ref{def:05} is used to highlight that the remote nodes belong to different partitions. In Fig.~\ref{fig:02}, we provide a pictorial representation for opposite remote sets. Clearly, vertices belonging to the same partition are remote \textit{per se}, as a consequence of the two-colorable graph state definition. The opposite remote set can be extended from vertices to set of vertices belonging to the same partition. In such a case, by considering a subset of vertices $A \subseteq P_i$ we formally define opposite remote set $\overline{N}_{\cup}(A)$ as the union of opposite remote set of each vertex in $A$, and $\overline{N}_{\cap}(A)$ as the intersection among the opposite remote sets of each vertex in $A$, i.e.:
\begin{align}
    \label{eq:03}
    \overline{N}_{\cup}(A) &\eqdef \bigcup_{v_i \in A} \overline{N}(v_i) \\
    \label{eq:04}
    \overline{N}_{\cap}(A) &\eqdef \bigcap_{v_i \in A} \overline{N}(v_i)
\end{align}

\begin{defin}[\textbf{Star vertex}]
    \label{def:06}
    Given a two-colorable graph $G = (P_1, P_2,E)$, the vertex $v_i$ belonging to partition $P_i$ is defined as star vertex if its neighborhood $N(v_i)$ coincides with the opposite partition $P_j\eqdef V \setminus P_i$, i.e.,:
    \begin{align}
        \label{eq:05}
         N(v_i) \eqdef \big\{ v_j \in V : (v_i,v_j) \in E \big \} \equiv V \setminus P_i \eqdef P_j.
    \end{align}
\end{defin}

In the following, for the sake of notation simplicity, we denoted with $S_1 \subseteq P_1$ and $S_2 \subseteq P_2$ the set of star vertices in the two partitions, and, accordingly, by denoting the remaining vertices, i.e. non-star vertices, in each partition as $V_1$ and $V_2$, we can adopt the following labeling for the two-colorable graph $G=(P_1,P_2,E)$:
\begin{align}
    \label{eq:06}
    P_1 = S_1 \cup V_1 \\
    \quad \text{ with } S_1 &= \{s^1_1,\cdots, s^{k_1}_1\} \wedge  V_1 = \{v^1_1,\cdots,v^{n_1}_1\} \nonumber \\
    \label{eq:07}
    P_2 = S_2 \cup V_2 \\
    \quad \text{ with } S_2 &= \{s^1_2,\cdots, s_2^{k_2} \} \wedge  V_2 = \{v^1_2,\cdots,v^{n_2}_2\} \nonumber
 \end{align}
with $|P_1| = n_1 + k_1$ and $|P_2| = n_2 + k_2 $.

\subsection{Problem Statement}
\label{sec:2.2}

Building on the concepts of remote $n$-Gability and Pairability introduced in Definitions~\ref{def:02}–\ref{def:03}, we now formally state the research problem.

\begin{prob} [Remote-VM problem]
    Given a graph state $\ket{G}$ distributed across the network nodes, we determine the number -- in the following referred to as \textit{volume} -- and position of $n$-qubit GHZ states and EPR pairs from $\ket{G}$, extractable among remote nodes, by using only single-qubit Clifford operations, single-qubit Pauli measurements and classical communication.
\end{prob}

As aforementioned, this problem is NP-complete, as the task of determining whether a specified set of Bell pairs (Bell-VM) or a specified GHZ (GHZ-VM) can be extracted from a graph state is already NP-complete~\cite{DahHelWeh-18,DahHelWeh-20}. The complexity of the Remote-VM problem is further exacerbated by constraining the entanglement resources to be extracted among remote nodes.

To address this challenge, we extend two sufficient conditions in \cite{CheCacCal-25} that enable the extraction of a quantifiable number of EPR pairs and $n$-qubit GHZ states among remote nodes from any given two-colorable graph state. 
In addition, we introduce a polynomial-time algorithm to further enhance the extractable volumes of both remote Pairability and remote $n$-Gability, applicable to arbitrary two-colorable graph states.

\begin{algorithm}[!t]
\caption{\textbf{\texttt{Remote Extraction}}{$(G, n)$}}
\label{alg:01} 
\begin{algorithmic}[1]
    \Require two-colorable graph  $G=(P_1, P_2, E)$
    \Ensure $\hat{A}_g, r_g(n)$
    
    \vskip 0.3em
    \Statex \hspace*{-\algorithmicindent} \rotatebox[origin=c]{-90}{\scalebox{0.8}{$\triangle$}}
    \textit{Step 1: Ensure star vertex exist in each partition}
    \vskip 0.3em
    
    \For{$(P, V)$ in $[(P_1, V_1), (P_2, V_2)]$}:
        \If{$P == V$}: \Comment{\textit{Partition $P$ lacks star vertex}}
            \State $v_i \gets$ random.choice($P$) 
            \State $G \gets G \setminus \overline{N}(v_i)$ 
        \EndIf  \Comment{\textit{Updated $v_i$ as star vertex in partition $P$}}
    \EndFor
    
    \vskip 0.3em
    \Statex \hspace*{-\algorithmicindent} \rotatebox[origin=c]{-90}{\scalebox{0.8}{$\triangle$}} \textit{Step 2: Drive initial $\tilde{r}_g(n) = |\tilde{A}_g|=\max\{|A_g|,|B_g|\}$ by random choosing $A_g, B_g$ in one partition}
    \vskip 0.3em
    
    \State $A_g \gets$ random.choice ($A_g \subseteq V_1:$ $A_g$ \text{satisfy} \eqref{eq:08}  \text{in} $G$)
    \State $B_g \gets$ random.choice ($B_g \subseteq V_1:$ $B_g$ \text{satisfy} \eqref{eq:09} \text{in} $G$)
    \If{$|B_g| > |A_g|$}
        \State $G, \tilde{A}_g \gets G \setminus \overline{N}_{\cap }(B_g), B_g$
    \Else
        \State $G, \tilde{A}_g \gets G, A_g$
    \EndIf

    \vskip 0.3em
    \Statex \hspace*{-\algorithmicindent} \rotatebox[origin=c]{-90}{\scalebox{0.8}{$\triangle$}}
    \textit{Step 3: Obtain final $r_g(n)= |\hat{A}_g|$ by updating $\tilde{A}_g$ to $\hat{A}_g$}
    \vskip 0.3em
    \State $\texttt{A}, \texttt{\={A}2A} \gets$ \Call{FindA}{$G, n, \tilde{A}_g$}
    \While{$\texttt{A} \neq \emptyset$}
        \If{$\exists v_i \in \texttt{A} \text{ with } |\texttt{A}(v_i)| = 0$}  
            \State $\tilde{A}_g \gets \tilde{A}_g \cup \{ v_i \}$ \Comment{\textit{Add $v_i$ to $\tilde{A}_g$}}
        \Else
            \State Select any $v_i \in \texttt{A}$
            \State $v_j \gets \texttt{\={A}2A}(v_i)$ \textbf{if} $|\texttt{\={A}2A}(v_i)| = 1$ \textbf{else} $\emptyset$
            \State $G \gets G \setminus (\overline{N}({v_i}) \cap \overline{N}_{\cup}({\tilde{A}_g}) \setminus \overline{N}(v_j))\setminus \{v_j\}$
            \State $\tilde{A}_g \gets (\tilde{A}_g \cup \{ v_i \}) \setminus \{ v_j \}$          
        \EndIf
        \State $\texttt{A}, \texttt{\={A}2A} \gets$ \Call{FindA}{$G, n, \tilde{A}_g$}    
    \EndWhile
    \State $\hat{A}_g, r_g(n) \gets \tilde{A}_g, |\tilde{A}_g|$
    \State \Return $\hat{A}_g, r_g(n)$
\end{algorithmic}
\end{algorithm}

\floatname{algorithm}{Function}
\renewcommand{\thealgorithm}{1}
\begin{algorithm}[t]
\setstretch{1}    
\caption{\textbf{\texttt{FindA}}$( G, n, \tilde{A}_g )$}
\label{fun:x02}
    \begin{algorithmic}[1]
    \State $\bar{A} \gets \{ v_i \in V_i \setminus \tilde{A}_g : \left( |\overline{N}({v_i})| \geq (n-1) \right) \wedge \left( \overline{N}({v_i}) \not\subseteq \overline{N}_{\cup}({\tilde{A}_g}) \right)\}$ 
    \State $ \texttt{A}, \texttt{B2A}, \texttt{\={A}2A} \gets \emptyset, \emptyset, \emptyset$
    
    \For{$v_i \in \bar{A}$}
        \State $\texttt{\={A}2A}(v_i) \gets \{ v_j \in \tilde{A}_g : \overline{N}(v_j) \subseteq \overline{N}(v_i) \}$
        \State $\texttt{B2A}(v_i) \gets \{ v_j \in \tilde{A}_g : \overline{N}(v_j) \cap \overline{N}(v_i) \neq \emptyset \}$
    \EndFor
    
    \vskip 0.3em    
    \Statex \hspace*{-\algorithmicindent} \rotatebox[origin=c]{-90}{\scalebox{0.8}{$\triangle$}} \textit{Find $\texttt{A}(v_i)$ from $\texttt{B2A}(v_i)$, which can be combined with $v_i$ to form $B_g$, satisfying Lem.~\ref{lem:02}.}
    \vskip 0.3em

    \For{$v_i \in \texttt{B2A}$}
        \If{$|\texttt{\={A}2A}(v_i)| \leq 1$ \textbf{and} $\forall v_k \in (\texttt{B2A}(v_i) \setminus \texttt{\={A}2A}(v_i)), \{v_i, v_k\}$ satisfies \eqref{eq:08} or \eqref{eq:09} in $G$}        
            \State $\texttt{A}(v_i) \gets \texttt{B2A}(v_i)$ 
        \EndIf
    \EndFor

    \State \Return $\texttt{A}, \texttt{\={A}2A}$    
    \end{algorithmic}
\end{algorithm}

\section{Algorithm}
\label{sec:3}
Here, in Sec.~\ref{sec:3.2}, we first generalize for any two-colorable graph state, two sufficient conditions for remote entanglement resource extraction, proved in \cite{CheCacCal-25}. Then, we design a polynomial algorithm for the extraction of remote entanglement resources, by exploiting the aforementioned two conditions. Finally, in Sec.~\ref{sec:3.3}, we analyze the time complexity of the proposed algorithm.

\subsection{Remote Extraction Algorithm}
\label{sec:3.2}

\subsubsection{Extraction Conditions}
As the first work on remote extraction, \cite{CheCacCal-25} proposes a series of sufficient conditions for both remote Pairability and remote $n$-Gability, specifically focusing on a certain type of two-colorable graph. Building on \cite{CheCacCal-25}, we generalize these conditions and provide a broader framework. And the concise introduction to the updated extraction conditions for general two-colorable graph states is provided in Appendix~\ref{app:conditions}. We refer interested readers to Lemma~\ref{lem:01} and ~\ref{lem:02} in Appendix~\ref{app:conditions}. In summary, we present the following updates to the existing conditions introduced in~\cite{CheCacCal-25}.
\begin{itemize}
    \item Since an EPR pair can be viewed as a 2-qubit GHZ state, we further refine the existing conditions by focusing on remote $n$-Gability and applying the corresponding remote Pairability condition for $n=2$.
    \item To overcome the limitation of considering only a specific type of two-colorable graph, we introduce updated qualifications in the sufficient conditions.
\end{itemize}

\begin{figure*}[t]
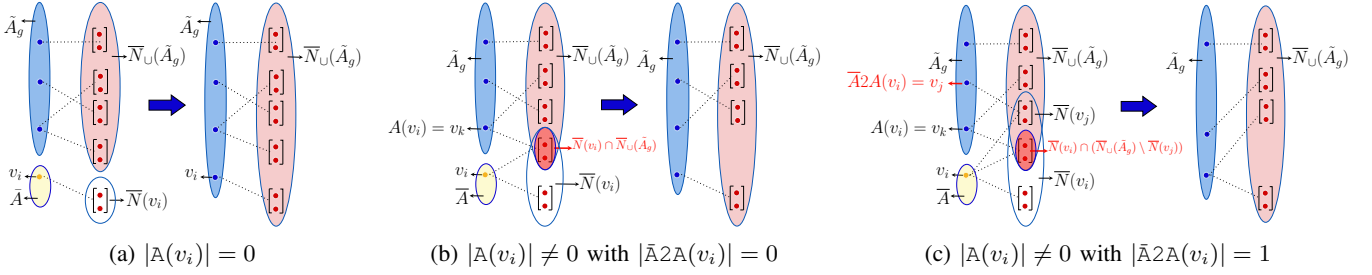

    \centering
    \begin{minipage}[b]{\textwidth}
    \centering
        \begin{minipage}[b]{0.275\textwidth}
            \centering\resizebox{\textwidth}{!}{
        \input{Figures/expandAa}
    }
            \subcaption{$|\texttt{A}(v_i)|= 0$}
    	   \label{fig:04.1}
        \end{minipage}
        \begin{minipage}[b]{0.33\textwidth}
            \centering\resizebox{\textwidth}{!}{
        \input{Figures/expandAb}
    }
            \subcaption{$|\texttt{A}(v_i)| \neq 0$ with $|\texttt{\={A}2A}(v_i)|= 0$}
    	   \label{fig:04.2}
        \end{minipage}
        \begin{minipage}[b]{0.38\textwidth}
            \centering\resizebox{\textwidth}{!}{
        \input{Figures/expandAc}
    }
            \subcaption{$|\texttt{A}(v_i)| \neq 0$ with $|\texttt{\={A}2A}(v_i)|= 1$}
    	   \label{fig:04.3}
        \end{minipage}
    \end{minipage}
    \caption{Pictorial representation for the step 3 (Line 14-25) in Algorithm~\ref{alg:01}. A preliminary set $\tilde{A}_g$, denoted with light blue background while $\overline{N}_{\cup}(\tilde{A}_g)$ is denoted with light red background. After \texttt{\textsc{FindA}}, it finds the set $\bar{A}$, for which the union of its opposite remote sets is not contained in $\overline{N}_{\cup}(A_g)$, denoted with light yellow background. Besides, it also computes corresponding $\texttt{\={A}2A}$ and $\texttt{A}$. Specifically, $\bar{A}2A(v_i)$, for which the opposite remote sets is contained in $\overline{N}(v_i)$. And  $A(v_i)$ can be combined with $v_i$ to form $A_g$ or $B_g$, satisfying~\eqref{eq:08} or~\eqref{eq:09}.
    Once $A \neq \emptyset$, it enters \texttt{\textsc{While}} loop in line 15-25.
    \textbf{(a)} During \texttt{\textsc{While}} loop, $v_i$ with $|\texttt{A}(v_i)|=0$ is firstly found in line 16. Then $v_i$ is directly added into  $\tilde{A}_g$. 
    \textbf{(b)} Once no $v_i$ with $|\texttt{A}(v_i)|=0$ exists, $v_i$ with $|\texttt{\={A}2A}(v_i)|=0$ can be found. Namely, it does not exist a node in $A_g$ whose opposite remote set is included in $\bar{N}(v_i)$. By removing $\overline{N}({v_i}) \cap \overline{N}_{\cup}(\tilde{A}_g)$, denoted with dark red background, $v_i$ is added into  $\tilde{A}_g$.
    \textbf{(c)} Once no $v_i$ with $|\texttt{A}(v_i)|=0$ exists, $v_i$ with $|\texttt{\={A}2A}(v_i)|=1$ can be found too. Namely, it exist only one node $\texttt{\={A}2A}(v_i)=v_j$, in $\tilde{A}_g$ whose opposite remote set is included in $\bar{N}(v_i)$. By removing $\overline{N}({v_i}) \cap \left( \overline{N}_{\cup}({\tilde{A}_g}) \setminus \overline{N}(v_j) \right)$, denoted with dark red background, $v_i$ is replaced by $v_j$ and added into $\tilde{A}_g$. Once \texttt{A} no longer exists, the \texttt{\textsc{while}} loop ends and updates $\tilde{A}_g$ to $\hat{A}_g$ in line 26.} 
    \label{fig:04}
    \hrulefill
\end{figure*}

\vspace{2mm}
\subsubsection{Algorithm Design}
Our polynomial-time algorithm for remote extraction in two-colorable graph state is described in Algorithm \ref{alg:01}, which is organized into three steps:
\begin{itemize}
    \item[-] Step 1. Let $G$ be the bipartite graph, if $G$ does not have star vertex in partition $P$, then it updates one vertex as new star vertex in partition $P$ (Line 1-6). More details refer to Cor.~\ref{cor:03} in App.~\ref{app:conditions}.
    
    \item[-] Step 2. Initially, we randomly choose two set $A_g$ and $B_g$ in $G$. By comparing the cardinality of $A_g$ and $B_g$ at lines 7-13, if $|B_g| > |A_g|$, once removing $\overline{N}_{\cap}(B_g)$ from $G$, $B_g$ becomes the new $\tilde{A}_g$. Until this step, we can preliminarily derive initial extractable values, i.e., $|\tilde{A}_g|=\max\{|A_g|,|B_g|\}$. For the sake of notation simplicity, let us denote $|\tilde{A}_g|$ as $\tilde{r}_g(n)$ in the following, which serves as the initial extractable volume of remote $n$-Gability. Formally:
    \begin{equation}
        \label{eq:11}
        \tilde{r}_g(n)=|\tilde{A}_g|= \max\{|A_g|,|B_g|\}.
    \end{equation}
    
    \item[-] Step 3. Alg.~\ref{alg:01} employs the strategy that operates on $\tilde{A}_g$, guiding it through a stepwise expansion in line 14 - 26, to obtain final extractable volume. Specifically, we provide a pictorial illustration for the step 3 in Fig.~\ref{fig:04}. \\
    It firstly calls Function~\texttt{\textbf{FindA}} in line 14 to calculate $\texttt{A}, \texttt{\={A}2A}$ in $\bar{A}$.  $\bar{A}$, for which the union of its opposite remote sets is not contained in $\overline{N}_{\cup}(A_g)$. Specifically, as stated in line 4 in~\texttt{FindA},  $\bar{A}2A(v_i)$, for which the opposite remote sets is contained in $\overline{N}(v_i)$. Besides, as stated in line 8-10 in~\texttt{FindA},  $A(v_i)$ can be combined with $v_i$ to form $B_g$, satisfying~\eqref{eq:09}.\\
    Then it follows a stepwise incremental approach within the \texttt{\textsc{while}} loop (lines 15–25), searching for a better vertex $v_i$ in $\bar{A}$. Based on its $|A(v_i)|$ and $|\bar{A}2A(v_i)|$, it either replaces a weaker vertex $\bar{A}2A(v_i)$ in $A_g$ or directly adds $v_i$ to $A_g$. Specifically, Fig.~\ref{fig:04} provides a conceptual overview of this coin process.

     Once \texttt{A} no longer exists, the \texttt{\textsc{while}} loop ends and updates $\tilde{A}_g$ to $\hat{A}_g$ in line 26. The cardinality, $r_g(n)$, of output $\hat{A}_g$ from Alg.~\ref{alg:01} is the final extractable volume for remote $n$-Gability. Based on above description, we can drive that:
    \begin{equation}      
    \label{eq:12}
         r_g(n) =|\hat{A}_g| \geq \, |\tilde{A}_g| =\max\{|A_g|,|B_g|\} =\tilde{r}_g(n).
    \end{equation}
    
\end{itemize}
 Hence, Alg.~\ref{alg:01} obtains the extractable values for both remote $n$-Gability and remote Pairability with $n=2$.
 
\subsection{Algorithm Complexity Analysis}
\label{sec:3.3}

Here, we analyze the complexity of Algorithm \ref{alg:01}. Let us assume, without any loss in generality, that the two-colorable graph $G=(P_1, P_2, E)$ is characterized by having $|P_1| \leq |P_2|$, with $P_1$ and $P_2$ defined in~\eqref{eq:06} and \eqref{eq:07}. We have the following attractive properties.

\begin{theo}
\label{theo:01}
For any two-colorable graph state $\ket{G}$, with corresponding graph $G=(P_1, P_2,E)$, Algorithm ~\ref{alg:01} determines the extractable volume $r_g(n)$ of $n$-qubit GHZ states among remote nodes and the extractable volume $r_g(2)$ of EPR pairs among remote nodes with time complexity $O(|P_1|^3*|P_2|)$.
    \begin{IEEEproof}
        Please refer to App.~\ref{app:theo:01}.   
    \end{IEEEproof}
\end{theo}

\begin{remark}
    Remarkably, our algorithm leverages graph theory and uses only single-qubit Clifford operations, single-qubit Pauli measurement and classical communication (LC + LPM + CC). Theorem~\ref{theo:01} shows that our algorithm can quickly determine the extractable volume. Furthermore, the final volume, obtained through successive updates, represents an enhanced extractable value, as shown in~\eqref{eq:12}. We refer interested readers to Appendix~\ref{app:theo:01} for the proof.
\end{remark}

\section{Performance Evaluation}
\label{sec:4}

\begin{figure*}[t]
    \centering    \includegraphics[width=\textwidth]{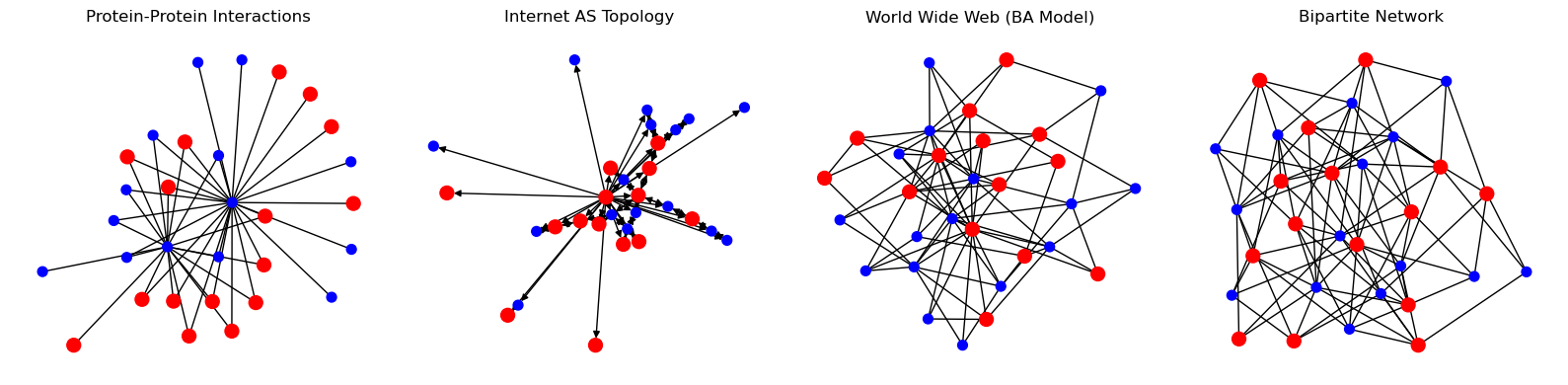}
    \caption{Complex artificial topologies used in evaluation. The graphs were generated using the NetworkX library.}
    \label{fig:x05}
    \hrulefill
\end{figure*}

In the following, we evaluate the extractable volume for both remote Pairability and remote $n$-Gability in general graph states.  The analysis is conducted on representative Internet-inspired artificial topologies,  as shown in Fig.~\ref{fig:x05}. Indeed, determining the exact extractable number in the optimal solution is NP-hard. In fact, solving this problem entails addressing the classical graph-theoretic challenge of counting independent sets, a well-known $\#$P-complete problem~\cite{DyeGre-00}. This complexity persists even for bipartite graphs, where counting independent sets is also $\#$P-complete~\cite{ProBal-83}. As a result, there exists no known algorithm, including those with exponential time complexity, that can reliably yield an optimal solution to the Remove-VM problem. Even in the bipartite case, where the theoretical upper bound of extractable volume is $\lceil\frac{N}{n}\rceil$, no known extraction strategy is capable of achieving this ideal in practice.

\begin{figure*}[t]
    \centering    \includegraphics[width=0.8\textwidth]{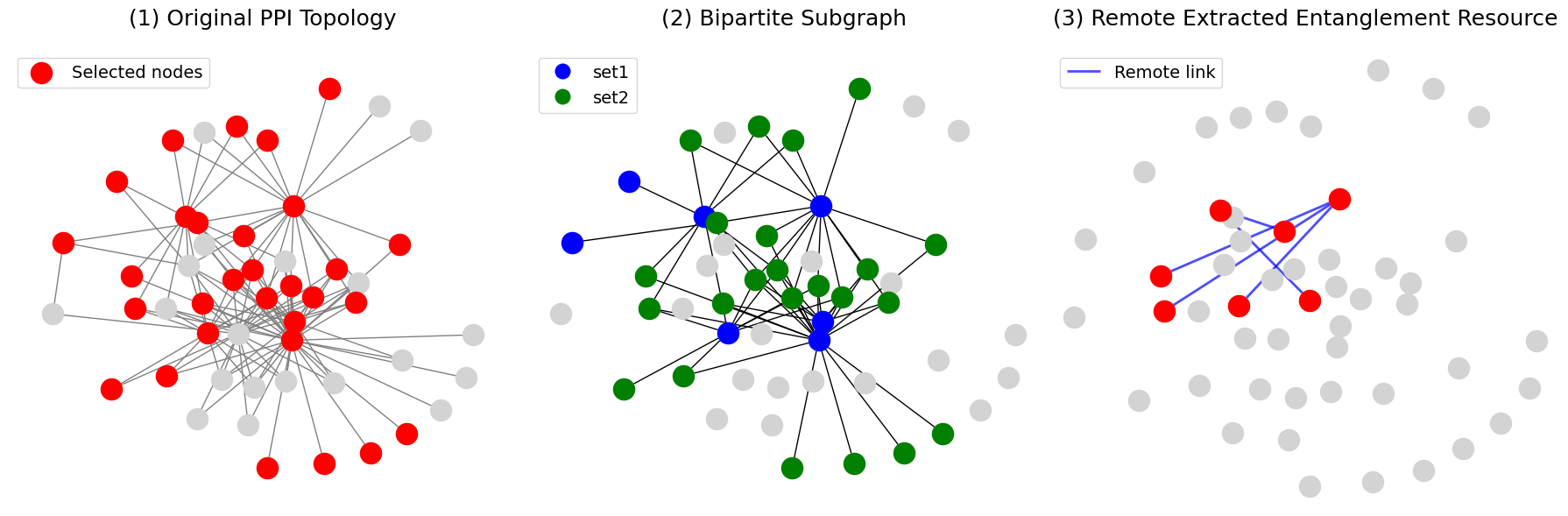}
    \caption{Pictorial illustration for Remote extraction from general graph state. 
    (1) The sample Protein-Protein Interactions (PPI) topology is generated by NetworkX library. (2) The extract bipartite subgraphs from PPI topology with fixed number of nodes. (3) The remote extracted entanglement resource, i.e., a $4$-qubit GHZ and a $3$-qubit GHZ, can be obtained by Algorithm~\ref{alg:01}.}
    \label{fig:xx06}
    \hrulefill
\end{figure*}

\begin{figure}[t]
    \centering    \includegraphics[width=0.5\textwidth]{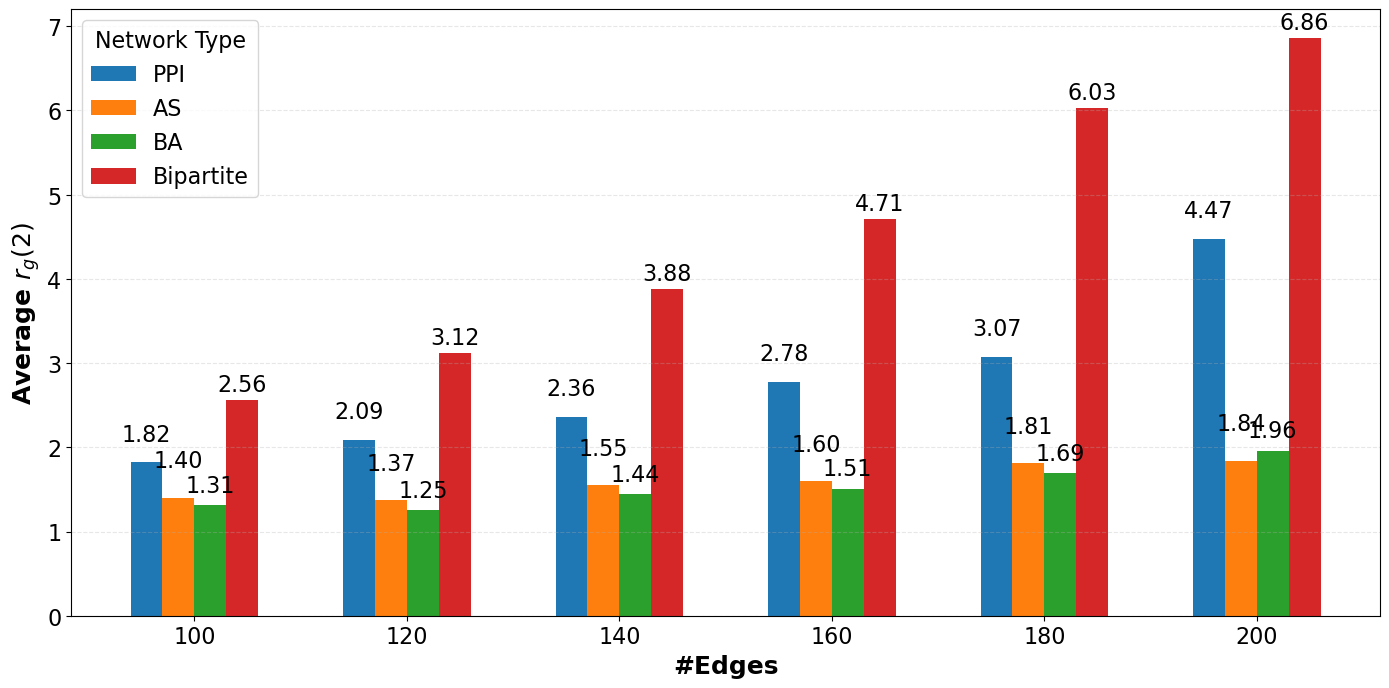}
    \caption{Remote Pairability Performance Analysis: Average extractable volume $r_g(2)$ in general 50-qubit graph state with Protein-Protein Interactions, AS Internet, World Wide Web Internet, and Bipartite network topology, respectively.}
    \label{fig:x06}
    \hrulefill
\end{figure}

\begin{figure}[t]
    \centering    \includegraphics[width=0.5\textwidth]{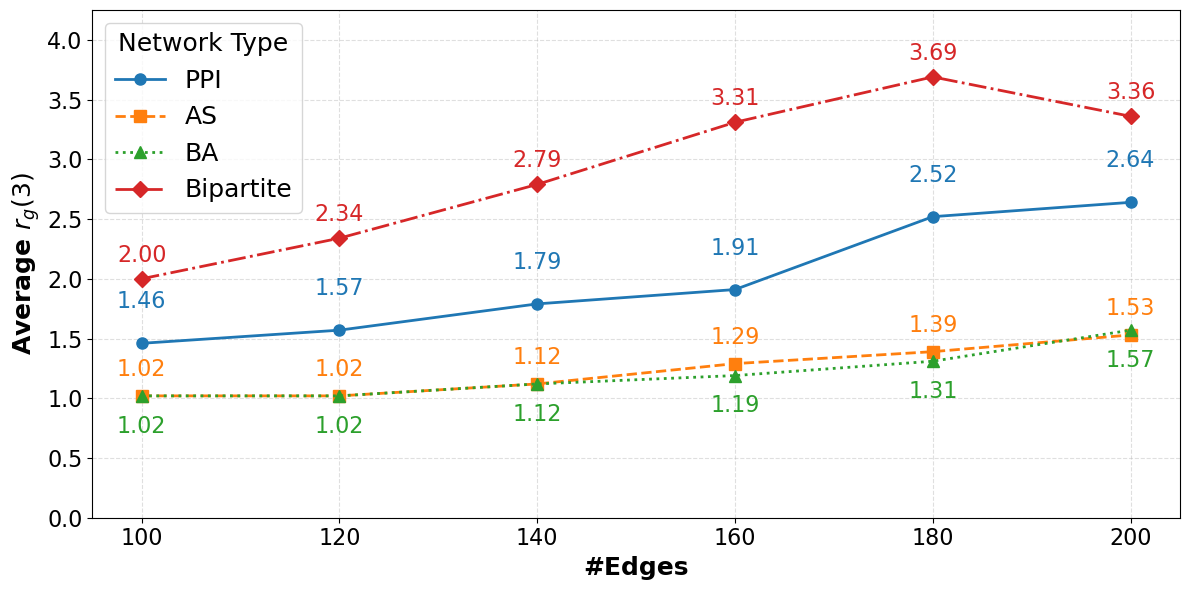}
    \caption{Remote 3-Gability Performance Analysis: Average extractable volume $r_g(3)$ in general 50-qubit graph state with Protein-Protein Interactions, AS Internet, World Wide Web Internet, and Bipartite network topology, respectively.}
    \label{fig:x07}
    \hrulefill
\end{figure}

\subsection{General Graph State Performance}
\label{sec:4.1}
We firstly evaluate the extractable volume $r_g(2), r_g(n)$ in general graph state against different graph structures. 
To better reflect the tested graph structures expected in future quantum networks, we selected four representative Internet topologies, as shown in Fig.~\ref{fig:x05}, to serve as artificial topologies for tested graph states. For each of these topologies, we conduct evaluations on two key metrics: remote pairability and remote $n$-Gability.

To ensure a fair and consistent evaluation across different topologies, we fix the total number of nodes at 50 and use the NetworkX library to generate graph instances by varying the number of edges $m$. However, as the original Internet topologies are not necessarily bipartite, Algorithm~\ref{alg:01} cannot be directly applied to compute their extractable volumes. To address this, we extract bipartite subgraphs from each 50-node topology by selecting 30 nodes that form a bipartite graph as shown in Fig.~\ref{fig:xx06}. To ensure statistical reliability, we performed 100 experiments to generate random graphs for each edge number scenario. This allows for a fair comparison across topologies under consistent structural conditions.

To evaluate the remote Pairability for general graph states, we compute the average for extractable values $r_g(2)$, as given in~\eqref{eq:12}. The Performance is shown in Fig.~\ref{fig:x06}. We observe that the extractable volume can be successfully determined for each type of Internet topology. Moreover, as the number of edges increases -- i.e., as the network topology becomes denser -- the extractable volume exhibits an approximately linear growth trend.


For remote n-Gability, we evaluate the extractable volume for 3-qubit GHZ states, i.e., $r_g(3)$, as a representative case. The results are shown in Fig.~\ref{fig:x07}. Similar to the remote pairability case, we are able to determine the extractable volume for all tested Internet topologies. In general, the extractable volume increases roughly linearly with edge density. However, we also observe a performance drop in extremely dense bipartite network structures, where the extractable volume does not continue to increase and may slightly decline.

\subsection{General Two-colorable Graph State Performance}
\label{sec:4.2}

\begin{figure*}[t]
    \centering
    \includegraphics[width=0.8\textwidth]{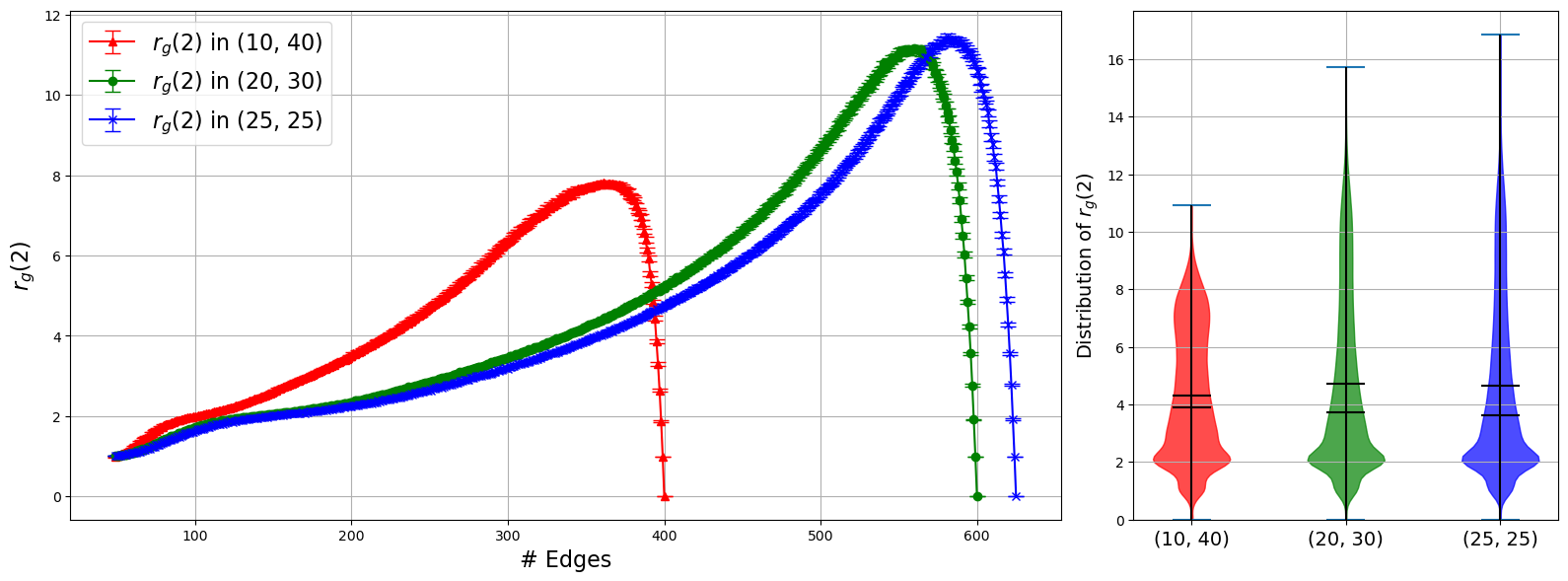}
    \caption{Remote Pairability Performance Analysis: $95\%$ confidence interval  and distribution of extractable volume $r_g(2)$ in two-colorable graph state with partitions (10, 40), (20, 30), (25, 25), respectively.}
    \label{fig:05}
    \hrulefill
\end{figure*}

\begin{figure*}[t]
    \centering
    \begin{subfigure}[b]{\textwidth}
        \begin{subfigure}[t]{0.325\textwidth}
            \centering
            \includegraphics[width=\textwidth]{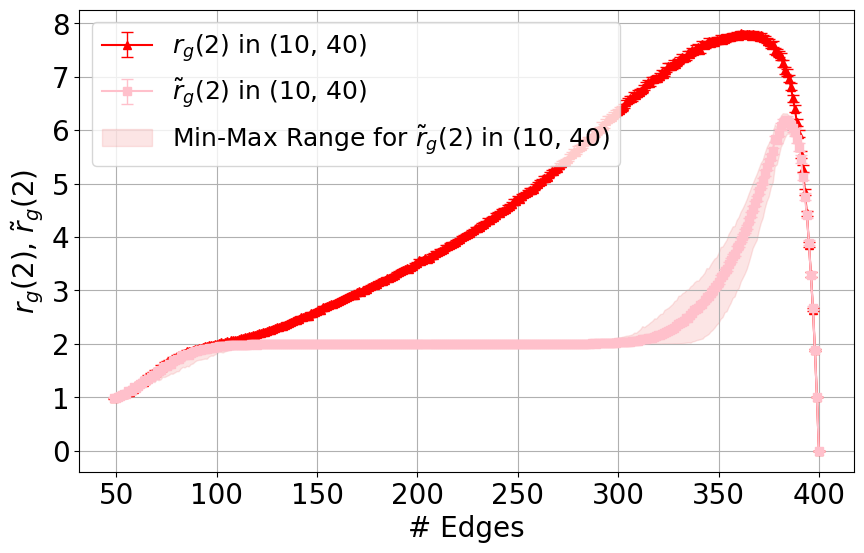}
            \subcaption{Graphs with partitions $(10, 40)$.}
            \label{fig:06.a}
        \end{subfigure}
        \hspace{3pt}
        \begin{subfigure}[t]{0.325\textwidth}
            \centering
            \includegraphics[width=\textwidth]{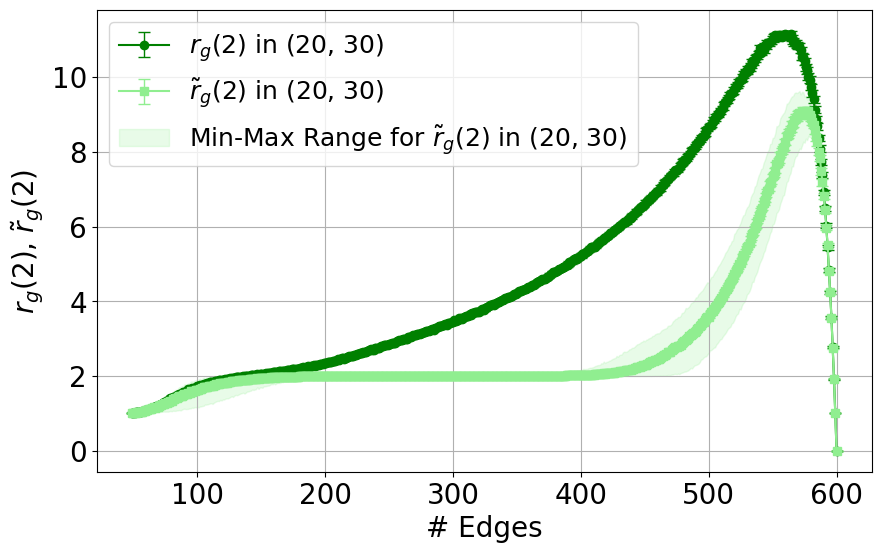}
            \subcaption{Graphs with partitions  $(20, 30)$.}
            \label{fig:06.b}
        \end{subfigure}
        \hspace{3pt}
        \begin{subfigure}[t]{0.325\textwidth}
            \centering
            \includegraphics[width=\textwidth]{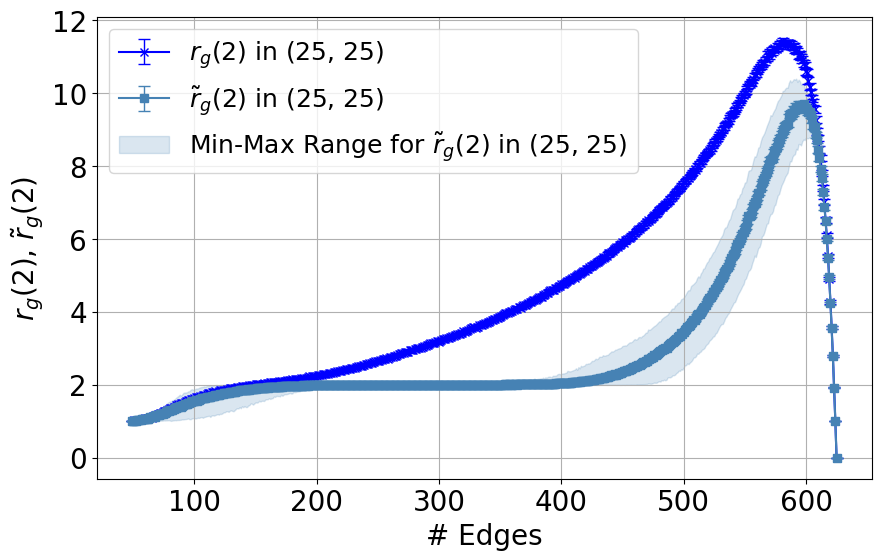}
            \subcaption{Graphs with partitions $(25, 25)$.}
            \label{fig:06.c}
        \end{subfigure}
    \end{subfigure}
    \caption{Algorithm Power Analysis: $95\%$ confidence interval for the final extractable value $r_g(2)$ against the preliminary extractable value $\tilde{r}_g(2)$. The figure also shows min-max range of $\tilde{r}_g(2)$ for each configuration of graph state.}
    \label{fig:06}
    \hrulefill
\end{figure*}

In the following, we conducted a comprehensive evaluation of general two-colorable graph states under various graph structures. An explicit setup process is provided, along with a detailed analysis of remote pairability and remote $n$-Gability.

\subsubsection{\textbf{Setup}}
\label{sec:4.2.1}

We evaluate the extractable values against different bipartite graph structures by randomly varying the number of edges $m$, while keeping the total number of nodes constant and equal to $50$. This allows for a fair comparison across various graph instances.
Furthermore, for the sake of generality, we distribute the nodes in different ways: one approach allocates nodes unequally across partitions, while the other ensures an equal number of nodes in each partition. More into details, we consider graphs with partitions ($P_1, P_2$) having sizes (10, 40), (20, 30), (25, 25), respectively. 

Accordingly, to Definition \ref{def:04} of two-colorable graph, we then randomly distribute the $m$ edges between the two partitions, thereby varying the graph's structure. However, it is worthwhile to note that for being adherent to the definitions of connected two-colorable graph, the number of edges has to satisfy some conditions, as highlighted in the following.\\
Let us suppose there are $m$ edges in a two-colorable graph $G=(P_1, P_2,E)$ and let us assume, without loss of generality, $|P_1| \leq |P_2|$, with $P_1$ and $P_2$ defined in~\eqref{eq:06} and \eqref{eq:07}. For a connected two-colorable graph $G$, the number of edges $m$ in $G$ satisfies the following inequality:
\begin{equation}
    \label{eq:13}
    (|P_1|+|P_2|-1) \leq m \leq |P_1|*|P_2|.
\end{equation}
When the number of edges in a bipartite graph exceeds the right-term in inequality \eqref{eq:13}, the graph no longer maintains the characteristics of two-colorable. While the number of edges in a bipartite graph unachieves the left-term in inequality \eqref{eq:13}, the graph can not be a connected two-colorable graph. To ensure statistical reliability, we performed 1000 experiments to generate random graphs for each edge number scenario.

\begin{figure*}[!t]
    \centering    \includegraphics[width=\textwidth]{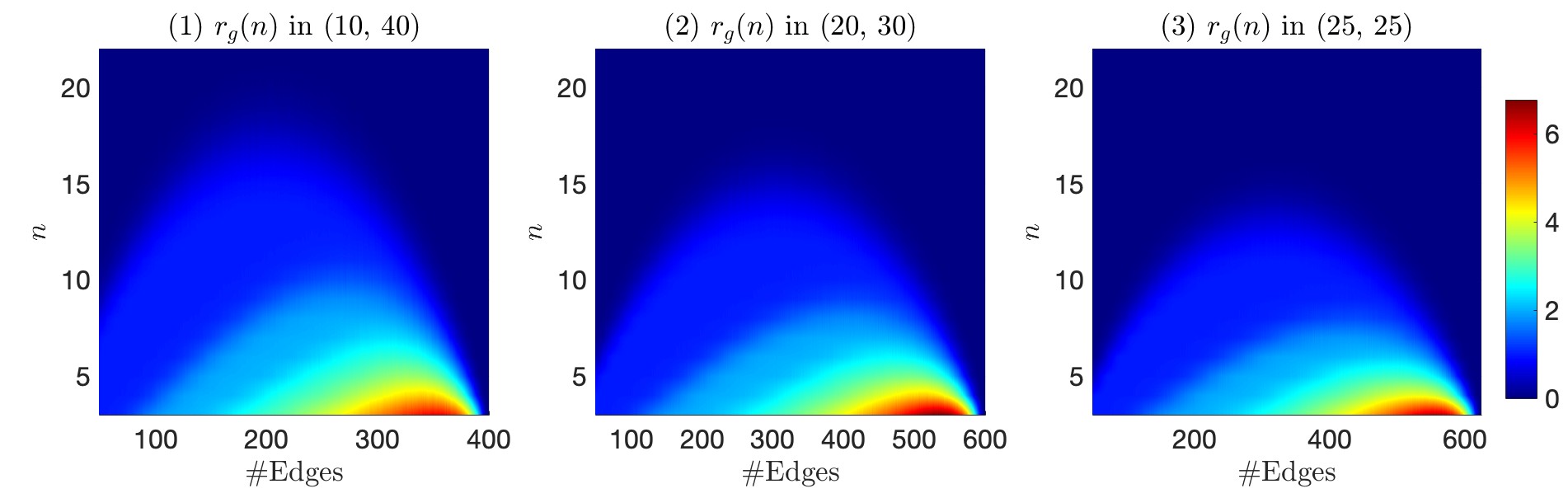}
    \caption{Remote $n$-Gability Performance Analysis: Average extractable volume $r_g(n)$ in graph states with Partitions $(10, 40)$, $(20, 30)$, $(25, 25)$.}
    \label{fig:07}
    \hrulefill
\end{figure*}

\subsubsection{\textbf{Remote Pairability Performance Analysis}}
\label{sec:4.2.2}
To evaluate the remote Pairability for general two-colorable graph states, we compute the $95\%$ confidence interval for extractable values $r_g(2)$, as given in~\eqref{eq:12}.

From the left part of Fig.~\ref{fig:05}, we observe an intriguing contrast in the performance of $r_g(2)$ across the three different partitions size of graph states. Specifically, the (25, 25) configuration exhibits the highest 
$r_g(2)$ values (11-12), indicating stronger or more pronounced extractable properties in balanced structures. The (20, 30) configuration shows intermediate values (10-11), while the (10, 40) configuration yields the lowest values (7–8), reflecting the impact of structural imbalance on the metric. Despite differences among the configurations, they exhibit a consistent trend in the behavior of 
$r_g(2)$ as the number of edges increases.

In the right part of Fig.~\ref{fig:05}, we also present the distribution of $r_g(2)$ across three graph configurations: (10, 40), (20, 30), and (25, 25). The red group (10, 40) exhibits the greatest variability with a bimodal trend; the green group (20, 30) displays a more balanced distribution; and the blue group (25, 25) is the most concentrated and stable.  Overall, the (25, 25) configuration demonstrates the greatest consistency, whereas the (10, 40) configuration shows the most pronounced fluctuations.

To demonstrate the effectiveness of our algorithm, in Fig.~\ref{fig:06} we also compare the preliminary extractable value $\tilde{r}_g(2)$, as given in~\eqref{eq:11}, with the final extractable value $r_g(2)$, derived using the optimization algorithm~\ref{alg:01}. Excluding the initial phase with sparse graphs and the final phase with highly dense graphs, the performance of 
$\tilde{r}_g(2)$ across all three configurations shows a dramatic improvement during the intermediate stages.

Related to the last observation, we further stress that the comparison between our bounds and existing literature is not fair, since our work is the first one, to the best of our knowledge, focusing on remote Pairability rather then on plain Pairability. More into details, regarding the Pariability, existing works, such as~\cite{Zha-25, BraShaSze-22}, propose algorithms to determine whether subsets of Bell pairs can be extracted from graph states with specific structures, such as rings, lines, and trees. Indeed,~\cite{Zha-25} provides conditions for extracting two EPR pairs from these structures, but they do not ensure remote extraction. Similarly,~\cite{BraShaSze-22} presents a 2-pairable 10-qubit graph state based on a ``wheel graph'' through exhaustive numerical evaluations of all the permutations of Pauli measurements on the qubits, without ensuring again remote extractions. Differently, with the same structure, our results assure one remote EPR extraction, through a significantly more constructive approach. 


\subsubsection{\textbf{Remote $n$-Gability Performance Analysis}}
\label{sec:4.2.3}

To evaluate the extractable values for remote Gability in two-colorable graph state, we compute the average for $r_g(n)$, which given in \eqref{eq:12}.

Fig.~\ref{fig:07} validates the $n$-Gability analysis, for each configuration of bipartite graph state and against not only the number $m$ of edges but also against the mass $n$ of the extracted GHZ states. As shown in Fig.~\ref{fig:07} the proposed approach  generally allows for the extraction of at least one GHZ state with a mass ranging from 3 to 17 among remote nodes. This implies that for a given graph state, one can typically extract a GHZ state of significant size among distant parties. 
Notably, when we consider $\ket{GHZ}_3$, we observe that the $r_g(3)$ surpasses 6 for each configuration  of graph state. This suggests that our approach facilitates the formation of small-scale GHZ states, i.e., of small subnets that can be exploited by entanglement-based protocols. 

Furthermore, we stress that existing studies focus on maximizing the mass of a single GHZ state, by limiting so the volume to be equal to one. For a graph state $\ket{G}$ with bounded rank-width, in~\cite{Oum-08} a poly-time algorithm determines whether a GHZ state can be extracted using local Clifford operations and Z-measurements, providing the required operation sequence. Similarly,~\cite{JonHahTch-24} demonstrates the extraction of GHZ states with masses from 4 to 11 starting from linear cluster states of up to 19 qubits on the IBMQ Montreal quantum device.
By accounting for the above, compared to existing studies, our results not only demonstrate the extraction of GHZ states with significantly larger
masses ranging from 3 to 17, but also ensure the extraction of a considerable volume of 3-qubit $\ket{GHZ}$ states. This showcases the versatility of our method, enabling both large and small-scale GHZ states, and providing a more scalable and efficient approach for quantum networks.



\begin{appendices}
\section{Extraction Conditions}
\label{app:conditions}
We propose two essential sufficient conditions for remote $n$-Gability in two-colorable graph states 
$\ket{G}$, where remote Pairability is treated as the special case with $n=2$. For a more comprehensive background, we recommend that readers refer to~\cite{CheCacCal-25}.

\begin{lem}[\textbf{Remote $n$-Gability: Condition I}]
    \label{lem:01}
    Let $\ket{G}$ be a two-colorable graph state, with corresponding graph $G=(P_1, P_2, E)$. A sufficient condition for concurrently extracting $\dot{r}_g(n)$ GHZ states, each involving $n$ qubits, is that at least $\dot{r}_g(n)$ vertices in one partition have pairwise disjoint opposite remote sets of cardinality at least $n - 1$, and that there exists at least one star vertex in each partition. Formally: 
    \begin{align}
        \label{eq:08} 
        & \exists S_1, S_2 \neq \emptyset, \exists A_g \subseteq V_i, \text{with}\, |A_g|=\dot{r}_g(n): \\
        &\nonumber |\overline{N}(v_i) | \geq n - 1, \forall v_i \in A_g  \; \; \wedge \\
        &\nonumber \overline{N}(v_i) \cap \overline{N}(v_j) \equiv \emptyset, 
       \forall v_i, v_j \in A_g, v_i \neq v_j.  
    \end{align}
    \begin{IEEEproof} 
    Please refer to the proof of Lemma 3 in~\cite{CheCacCal-25}. 
    \end{IEEEproof}
\end{lem}

Since an EPR pair can be regarded as a degenerate case of a GHZ state involving two qubits, the sufficient condition in Lemma~\ref{lem:01} with $n = 2$ applies directly to the concurrent extraction of $\dot{r}_g(2)$ EPR pairs. This result is formally stated in Corollary~\ref{cor:01}.

\begin{cor}[\textbf{Remote Pairability: Condition I}]
    \label{cor:01}
    Let $\ket{G}$ be a two-colorable graph state, with corresponding graph $G=(P_1, P_2, E)$, and let $A_g$ denote the set with $n=2$ defined in \eqref{eq:08}. A sufficient condition for concurrently extracting $\dot{r}_g(2)$ EPR pairs among remote nodes is that at least $\dot{r}_g(2)$ vertices in $A_g$, and that there exists at least one star vertex in each partition. 
\end{cor}

\begin{lem}[\textbf{Remote $n$-Gability: Condition II}]
    \label{lem:02}
    Let $\ket{G}$ be a two-colorable graph state, with corresponding graph $G=(P_1, P_2, E)$. A sufficient condition for concurrently extracting $\ddot{r}_g(n)$ GHZ states, each involving $n$ qubits, is that at least $\ddot{r}_g(n)$ vertices in one partition have opposite remote sets that share only one unique intersection, with each opposite remote set remaining at least $(n-1)$ vertices after excluding this intersection; and that there exists at least one star vertex in each partition. Formally:
    \begin{small}
    \begin{align}
        \label{eq:09}
        & \exists S_1, S_2 \neq \emptyset, \exists! \overline{N}_{\cap}(B_g),\, \text{with} \, B_g \subseteq V_i \,\text{and}\, |B_g|=\ddot{r}_g(n) : \\&\nonumber
         |\overline{N}(v_i) \setminus \overline{N}_{\cap}(B_g)| \geq n-1, \forall v_i \in B_g \; \wedge \\
        & \nonumber \left( \overline{N}(v_i) \setminus \overline{N}_{\cap}(B_g) \right) \cap \left( \overline{N}(v_j) \setminus \overline{N}_{\cap}(B_g) \right) \equiv \emptyset, \forall v_i, v_j \in B_g, v_i \neq v_j.
    \end{align}
    \end{small}
with $\overline{N}_{\cap}(B_g) \subset P_j \neq P_i$  defined in \eqref{eq:04} and denoting the unique intersection among the opposite remote sets of $B_g$.
    \begin{IEEEproof}
        Please refer to the proof of Lemma 4 in~\cite{CheCacCal-25}.   
    \end{IEEEproof}
\end{lem}

Similarly, the sufficient condition in Lemma~\ref{lem:02} with $n = 2$ applies directly to the concurrent extraction of $\ddot{r}_g(2)$ EPR pairs. This result is formally stated in Corollary~\ref{cor:02}.

\begin{cor}[\textbf{Remote Pairability: Condition II}]
    \label{cor:02}
    Let $\ket{G}$ be a two-colorable graph state, with corresponding graph $G=(P_1, P_2, E)$ and let $B_g$ denote the set with $n=2$ defined in \eqref{eq:09}. A sufficient condition for concurrently extracting $\ddot{r}_g(2)$ EPR pairs, is that at least $\ddot{r}_g(2)$ vertices in $B_g$; and that there exists at least one star vertex in each partition. 
\end{cor}

Given that only one partition or no partition in $\ket{G}$ contains star vertices, neither Lemmas~\ref{lem:01} nor \ref{lem:02} can be exploited to assess $n$-Gability and pairability. Here, we work toward such an issue by introducing additional graph manipulations, as formally defined in Cor.~\ref{cor:03}.

\begin{cor}
    \label{cor:03}
    Let $\ket{G}$ be a two-colorable graph state, with corresponding graph $G=(P_1, P_2, E)$, where $P_j=V_j$, implying that partition $P_j$ contains no star vertices. $G$ can be reduced to a graph $G'$, vertex minor of $G$, as follows:
    \begin{equation}
        \label{eq:10}
        G' = G \setminus \overline{N}(v^i_j)
    \end{equation}
    with $v^i_j$ denoting the new star vertex in partition $P_j$.
\end{cor}

\section{PROOF OF THEOREM~\ref{theo:01}}
\label{app:theo:01}

For initialization, it takes $O(|P_1|*|P_2|)$ time to ensure both partition with at least one star vertex in Line 1-6. 

For random.choice $A_g, B_g$ in Line 7-8, we firstly random generate one permutation of $V_1$ and then sequently check nodes in this permutation to group as $A_g, B_g$. This process should compute opposite remote sets for nodes in $P_1$, which in $O(|P_1|*|P_2|)$. Generates a random permutation and processes nodes, which is $O(|P_1|^2*|P_2|)$ in the worst case (due to nested loops and checks). If the cardinality of random $B_g$ is larger than random $A_g$, we remove the intersection of opposite remote sets for $B_g$, namely $\overline{N}_{\cap}(B_g)$, which needs $O(|P_1|*|P_2|)$ time to compute in the worst case ($|B_g| = |P_1|$). 

Once entering \texttt{\textbf{FindA}} in Line 14, firstly compute $\bar{A}$ in worst case, which involves opposite remote set calculations for nodes in $P_1$ with complex $O(|P_1|*|P_2|)$. For each $v_i \in \bar{A}2A$, it needs to check subset condition for each $v_j \in \tilde{A}_g$ with $O(|\tilde{A}_g|*|P_2|)$ time. In total, it needs $O(|\bar{A}|*|\tilde{A}_g|*|P_2|)$ time in worst time to compute $\bar{A}2A$. Similarly as $B2A$. To check condition in Line 8 in \texttt{\textbf{FindA}}, for each $v_i \in B2A$, it takes $O(|B2A(v_i)|*|P_2|)$ to check whether they can group as $A_g$ or $B_g$. In worst case, total for all $v_i$ it takes $O(|\bar{A}|*|P_1|*|P_2|) \approx O(P_1|^2*|P_2|)$ time with $B2A(v_i)\approx |\tilde{A}_g|\approx |P_1|$. 

After that, it enters the \texttt{While} loop in Line 15-25 in Alg.~\ref{alg:01}, each iteration involves two case, Case 1: $A(v_i) = \emptyset$ and Case 2: $\texttt{A}(v_i) \neq \emptyset$. 
\begin{itemize}
    \item In Case 1, it takes $O(\texttt{A})$ to find such $v_i$ and then recompute $\texttt{A}, \bar{A}2A$ via \texttt{\textbf{FindA}} with $O(P_1|^2*|P_2|)$ time.
    \item In Case 2, compute the opposite remote set of random $v_i$ with $O(|P_2|)$ time, and compute the union of opposite remote set of $\tilde{A}_g$, i.e., $\overline{N}_{\cup}(\tilde{A}_g)$ takes $O(|\tilde{A}_g|*|P_2|)$. If it exists $|\bar{A}2A(v_i)|=1$, we need to spend  $O(|P_2|)$ time to calculate the opposite remote set of $v_j$, which is $\bar{A}2A(v_i)$. Then perform a series of $O(1)$ operation.
\end{itemize}
After that, it still needs to recompute $\texttt{A}, \bar{A}2A$ via \texttt{\textbf{FindA}} with $O(P_1|^2*|P_2|)$ time. The \texttt{While} loop terminates when \texttt{A} is empty. In worst-case iterations, the loop can run $O(|P_1|)$ times, as each iteration either Case 1 or Case 2. Total complexity of the \texttt{While} loop is  $O(P_1|^3*|P_2|)$.

Based on above analysis, the complexity of Alg.~\ref{alg:01} is $O(P_1|^3*|P_2|)$, which is polynomial - time algorithm.

\end{appendices}

\bibliographystyle{IEEEtran}
\bibliography{biblio.bib}

@article{CacCalVan-20,
 title={When entanglement meets classical communications: Quantum teleportation for the quantum internet},
	author={Cacciapuoti, Angela Sara and Caleffi, Marcello and Van Meter, Rodney and Hanzo, Lajos},
	journal={IEEE Transactions on Communications},
	volume={68},
	number={6},
	pages={3808--3833},
	year={2020},
	publisher={IEEE},
	note = {invited paper},
}

@article{DahHelWeh-20,
  title={How to transform graph states using single-qubit operations: computational complexity and algorithms},
  author={Dahlberg, Axel and Helsen, Jonas and Wehner, Stephanie},
  journal={Quantum Science and Technology},
  volume={5},
  number={4},
  pages={045016},
  year={2020},
  publisher={IOP Publishing}
}

@article{DahHelWeh-18,
  title={Transforming graph states to Bell-pairs is NP-Complete},
  author={Dahlberg, Axel and Helsen, Jonas and Wehner, Stephanie},
  journal={Quantum},
  volume={4},
  pages={348},
  year={2020},
  publisher={Verein zur F{\"o}rderung des Open Access Publizierens in den Quantenwissenschaften}
}

@article{IllCalMan-22,
    title={Quantum Internet Protocol Stack: a Comprehensive Survey},
    author={Illiano, Jessica and Caleffi, Marcello and Manzalini, Antonio and Cacciapuoti, Angela Sara},
    journal={Computer Networks},
    volume={213},
    year={2022}
}

@article{PirDur-18,
  title={Modular architectures for quantum networks},
  author={Pirker, A and Walln{\"o}fer, J and D{\"u}r, W},
  journal={New Journal of Physics},
  volume={20},
  number={5},
  pages={053054},
  year={2018},
  publisher={IOP Publishing}
}

@article{BraShaSze-22,
  title={Generating $ k $ EPR-pairs from an n-party resource state},
  author={Bravyi, Sergey and Sharma, Yash and Szegedy, Mario and de Wolf, Ronald},
  journal={arXiv preprint arXiv:2211.06497},
  year={2022}
}

@article{CheIllCac-25,
      title={Entanglement-based artificial topology: Neighboring remote network nodes},
      author={Chen, Si-Yi and Illiano, Jessica and Cacciapuoti, Angela Sara and Caleffi, Marcello},
      journal={IEEE Open Journal of the Communications Society},
      year={2025},
      publisher={IEEE}
}

@article{JonHahTch-24,
  title={Extracting GHZ states from linear cluster states},
  author={de Jong, Jarn and Hahn, Frederik and Tcholtchev, Nikolay and Hauswirth, Manfred and Pappa, Anna},
  journal={Physical Review Research},
  volume={6},
  number={1},
  pages={013330},
  year={2024},
  publisher={APS}
}

@article{ProBal-83,
author = {Provan, J. Scott and Ball, Michael O.},
title = {The Complexity of Counting Cuts and of Computing the Probability that a Graph is Connected},
year = {1983},
issue_date = {Nov 1983},
publisher = {Society for Industrial and Applied Mathematics},
address = {USA},
volume = {12},
number = {4},
issn = {0097-5397},
url = {https://doi.org/10.1137/0212053},
doi = {10.1137/0212053},
journal = {SIAM J. Comput.},
month = {nov},
pages = {777–788},
numpages = {12}
}

@article{Zha-25,
  title={Bell pair extraction using graph foliage techniques},
  author={Zhang, Derek},
  journal={Journal of Mathematical Physics},
  volume={66},
  number={2},
  year={2025},
  publisher={AIP Publishing}
}

@article{Oum-08,
author = {Oum, Sang-Il},
title = {Approximating rank-width and clique-width quickly},
year = {2008},
issue_date = {November 2008},
publisher = {Association for Computing Machinery},
address = {New York, NY, USA},
volume = {5},
number = {1},
issn = {1549-6325},
doi = {10.1145/1435375.1435385},
journal = {ACM Trans. Algorithms},
month = dec,
articleno = {10},
numpages = {20}
}

@article{MazCalCac-24,
  title={Intra-QLAN Connectivity via Graph States: Beyond the Physical Topology},
  author={Mazza, Francesco and Caleffi, Marcello and Cacciapuoti, Angela Sara},
  journal={IEEE Transactions on Network Science and Engineering},
  year={2025},
  publisher={IEEE}
}

@article{DyeGre-00,
  title={On Markov chains for independent sets},
  author={Dyer, Martin and Greenhill, Catherine},
  journal={Journal of Algorithms},
  volume={35},
  number={1},
  pages={17--49},
  year={2000},
  publisher={Elsevier}
}

@article{CheCacCal-25,
      title={Entanglement-Enabled Connectivity Bounds for Quantum Network}, 
      author={SiYi Chen and Angela Sara Cacciapuoti and Marcello Caleffi},
      year={2025},
      number={submit/6359916},
      journal={arXiv},
      primaryClass={quant-ph} 
}

\end{document}